\documentclass{emulateapj}
\usepackage{apjfonts}
\usepackage{lscape}

\slugcomment{Submitted to ApJ 2006 February 17; accepted 2006 June 7}
\shorttitle{INTERSTELLAR ABSORPTION TOWARD THE PLEIADES}
\shortauthors{RITCHEY ET AL.}

\begin{document}
\title{The Nature of Interstellar Gas toward the Pleiades Revealed in 
Absorption Lines}
\author{A. M. Ritchey\altaffilmark{1}, 
M. Martinez\altaffilmark{1}$^,$\altaffilmark{2}, 
K. Pan\altaffilmark{3}$^,$\altaffilmark{4}, 
S. R. Federman\altaffilmark{1}$^,$\altaffilmark{5}, 
and D. L. Lambert\altaffilmark{6}}
\altaffiltext{1}{Department of Physics and Astronomy, University of Toledo, 
Toledo, OH 43606; aritchey@physics.utoledo.edu; steven.federman@utoledo.edu.}
\altaffiltext{2}{Department of Aeronautical and Astronautical Engineering, 
University of Washington, Seattle, WA 98195; marleen@u.washington.edu.}
\altaffiltext{3}{Department of Physics and Astronomy, Bowling Green State 
University, Bowling Green, OH 43403.}
\altaffiltext{4}{Apache Point Observatory, New Mexico State University, P.O. 
Box 59, Sunspot, NM 88349; kpan@apo.nmsu.edu.}
\altaffiltext{5}{Guest Observer, McDonald Observatory, University of Texas at 
Austin, Austin, TX 78712.}
\altaffiltext{6}{W.J. McDonald Observatory, University of Texas at Austin, 
Austin, TX 78712; dll@astro.as.utexas.edu.}

\begin{abstract}

We present high-resolution, high signal to noise absorption-line observations 
of CN, \ion{Ca}{2}, \ion{Ca}{1}, CH$^+$, and CH along twenty lines of sight 
toward members of the Pleiades. The acquired data enable the most detailed 
study to date of the interaction between cluster stars and the surrounding 
interstellar gas. Total equivalent widths are consistent with previous 
investigations except where weaker features are detected owing to our greater 
sensitivity. Mean $b$-values for the molecular species indicate that toward 
most of the Pleiades CH is associated with the production of CH$^+$ rather 
than CN. An analysis of radial velocities reveals a kinematic distinction 
between ionized atomic gas and molecular and neutral gas. Molecular components 
are found with velocities in the local standard of rest of either $\sim$ +7 km 
s$^{-1}$ or $\sim$ +9.5 km s$^{-1}$, with the higher-velocity components 
associated with the strongest absorption. Atomic gas traced by \ion{Ca}{2} 
shows a strong central component at $v_{\mathrm{\scriptstyle LSR}}$ $\sim$ +7 
km s$^{-1}$ exhibiting velocity gradients indicative of cloud-cluster 
interactions. Gas density estimates derived from measured CH/CH$^+$ column 
density ratios show good agreement with those inferred from H$_2$ rotational 
populations, yielding typical values of $n\sim50$ cm$^{-3}$. Our models do not 
include the important time-dependent effects on CH$^+$ formation which may 
ultimately be needed to extract physical conditions in these clouds.

\end{abstract}


\section{INTRODUCTION}

The interstellar medium (ISM) in the vicinity of the Pleiades is a rich 
environment for the study of processes that result from the interactions 
between stellar photons and the gas and dust clouds of interstellar space. The 
stars of this cluster were not formed out of the surrounding material visible 
as reflection nebulosity. Rather, the spatial association of the stars and the 
interstellar gas is the result of a chance encounter between the cluster and 
one or more approaching clouds (White 2003). Such collisions precipitate 
numerous radiative processes including the photoionization of atomic and 
molecular species, the photodissociation of molecules such as CH and H$_2$, 
and the photoelectric heating of diffuse gas by dust grains stimulated by 
ultraviolet radiation, all of which may help to explain some peculiarities of 
the ISM near the Pleiades.

Two such peculiarities are the great strength of CH$^+$ absorption lines 
observed toward many cluster members (White 1984\emph{a}) and the large amount 
of H$_2$ in rotationally excited states (Spitzer, Cochran, \& Hirshfeld 1974). 
While these anomalies may be related, a rigorous chemical model of CH$^+$ 
production in the Pleiades, and elsewhere in the ISM, remains to be developed. 
CH$^+$ is unable to form at the low temperature of diffuse clouds because the 
reaction leading to its formation, C$^+$ + H$_2$ $\to$ CH$^+$ + H, is 
endothermic with an activation energy of $\Delta E/k=4640$ K. Elitzur and 
Watson (1978, 1980) showed that the above reaction can produce sufficient 
amounts of CH$^+$ if the gas is heated by an interstellar shock. However, 
subsequent observations failed to detect a corresponding overabundance of OH 
molecules resulting from a similar endothermic reaction (Federman et al. 
1996\emph{a}). Several groups (Draine \& Katz 1986; Pineau des For\^ets et al. 
1986) employed magnetohydrodynamic (MHD) shocks to lessen the problem with the 
overproduction of OH, but even their models yielded OH column densities in 
excess of observations. The velocity shifts between CH and CH$^+$ absorption 
lines predicted by these shock models have also not been detected (e.g., 
Gredel, van Dishoeck, \& Black 1993). Alternate theories for a non-thermal 
origin of CH$^+$ have since been proposed. Federman et al. (1996\emph{b}) 
considered the motions of C$^+$ ions influenced by the passage of Alfv\'en 
waves against a static background of cold neutral gas. Their model predicts 
CH$^+$, CH, and OH column densities in line with observations, but requires an 
additional mechanism to account for observed HCO$^+$ abundances. More 
recently, non-equilibrium chemistry was investigated by Joulain et al. (1998) 
and Falgarone et al. (2005). These authors ascribe the transient heating of 
localized regions of the cold diffuse medium to intermittent bursts of 
turbulent dissipation by either MHD shocks or coherent small-scale vortices. 
Such models may be appropriate for many Galactic sight lines, yet the 
observational evidence in the Pleiades favors heating either by H$_2$ 
dissociation or by dust photoelectron emission, rather than shocks, as the 
agent responsible for CH$^+$ formation (White 1984\emph{b}).

\tabletypesize{\scriptsize}
\begin{deluxetable*}{lcccccccc}
\tablecolumns{9}
\tablewidth{0.8\textwidth}
\tablenum{1}
\tablecaption{Stellar and Observational Data}
\tablehead{\colhead{HD} & \colhead{Name} & \colhead{Type} & \colhead{$E(B-V)$} 
& \colhead{$B$} & \colhead{$\alpha$ [J2000]} & \colhead{$\delta$ [J2000]} & 
\colhead{$d$} & \colhead{$\tau_{exp}$} \\
\colhead{} & \colhead{} & \colhead{} & \colhead{(mag)} & \colhead{(mag)} & 
\colhead{($^h$, $^m$, $^s$)} & \colhead{($^{\circ}$, $^{\prime}$, 
$^{\prime\prime}$)} & \colhead{(pc)} & \colhead{(s)} }
\startdata
23288 & 16 Tau & B7 IV & \phs 0.10 & 5.41 & 03 44 48.22 & +24 17 22.1 & 103 
$\pm$ 11 & $3 \times 1200$ \\
23302 & 17 Tau & B6 III & \phs 0.05 & 3.61 & 03 44 52.54 & +24 06 48.0 & 114 
$\pm$ 12 & $2 \times \phn 600$ \\
23324 & 18 Tau & B8 V & \phs 0.05 & 5.58 & 03 45 09.74 & +24 50 21.3 & 113 
$\pm$ 11 & $2 \times 1800$ \\
23338 & 19 Tau & B6 IV & \phs 0.04 & 4.20 & 03 45 12.49 & +24 28 02.2 & 114 
$\pm$ 14 & $3 \times \phn 600$ \\
23408 & 20 Tau & B7 III & \phs 0.07 & 3.81 & 03 45 49.61 & +24 22 03.9 & 110 
$\pm$ 13 & $2 \times \phn 600$ \\
23410 & & A0 V & \phs 0.08 & 6.92 & 03 45 48.82 & +23 08 49.7 & 103 $\pm$ 11 & 
$4 \times 1800$ \\
23432 & 21 Tau & B8 V & \phs 0.07 & 5.73 & 03 45 54.48 & +24 33 16.2 & 119 
$\pm$ 13 & $2 \times 1800$ \\
23441 & 22 Tau & B9 V & \phs 0.06 & 6.42 & 03 46 02.90 & +24 31 40.4 & 109 
$\pm$ 11 & $2 \times 1800$ \\
23480 & 23 Tau & B6 IV & \phs 0.10 & 4.11 & 03 46 19.57 & +23 56 54.1 & 110 
$\pm$ 13 & $3 \times \phn 600$ \\
23512 & & A1 V & \phs 0.35 & 8.47 & 03 46 34.20 & +23 37 26.5 & $\ldots$ & $8 
\times 1800$ \\
23568 & & B9.5 V & \phs 0.07\tablenotemark{a} & 6.85 & 03 46 59.40 & +24 31 
12.4 & 150 $\pm$ 23 & $4 \times 1800$ \\
23629 & 24 Tau & A2 V & $-0.03$ & 6.30 & 03 47 21.04 & +24 06 58.6 & $\ldots$ 
& $3 \times 1800$ \\
23630 & $\eta$ Tau & B7 III & \phs 0.04 & 2.81 & 03 47 29.08 & +24 06 18.5 & 
113 $\pm$ 13 & $2 \times \phn 300$ \\
23753 & & B8 V & \phs 0.04 & 5.38 & 03 48 20.82 & +23 25 16.5 & 104 $\pm$ 10 & 
$2 \times 1200$ \\
23850 & 27 Tau & B8 III & \phs 0.04 & 3.54 & 03 49 09.74 & +24 03 12.3 & 117 
$\pm$ 14 & $2 \times \phn 600$ \\
23862 & 28 Tau & B8 Vp & \phs 0.03 & 4.97 & 03 49 11.22 & +24 08 12.2 & 119 
$\pm$ 12 & $3 \times \phn 900$ \\
23873 & & B9.5 V & \phs 0.02\tablenotemark{a} & 6.59 & 03 49 21.75 & +24 22 
51.4 & 125 $\pm$ 14 & $3 \times 1800$ \\
23923 & & B8 V & \phs 0.06\tablenotemark{a} & 6.12 & 03 49 43.53 & +23 42 42.7 
& 117 $\pm$ 13 & $2 \times 1800$ \\
23964 & & B9.5 Vp & \phs 0.11 & 6.80 & 03 49 58.06 & +23 50 55.3 & 159 $\pm$ 
39 & $4 \times 1800$ \\
24076 & & A2 V & \phs 0.03\tablenotemark{a} & 7.02 & 03 50 52.43 & +23 57 41.3 
& 102 $\pm$ 10 & $4 \times 1800$ \\
\enddata
\tablenotetext{a}{Reddening values for these stars were updated in White 
(2003).}
\end{deluxetable*}

\begin{figure}
\centering
\includegraphics[width=0.4\textwidth]{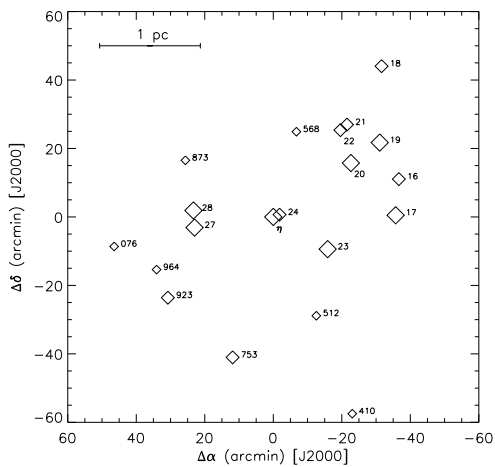}
\caption[Distribution of observed Pleiades stars.]{Distribution of observed 
Pleiades stars in relative equatorial coordinates centered on $\eta$ Tau 
[$\alpha(2000)=03^h47^m29^s.08$, 
$\delta(2000)=+24^{\circ}06^{\prime}18^{\prime\prime}.5$]. The label next to 
each symbol is either the last three digits of the HD number or the Flamsteed 
or Bayer designation. Decreasing symbol size denotes three magnitude ranges: 
$B < 5.0$, $5.0 < B < 6.5$, and $6.5 < B$. This key is used in Figures 6, 7, 
and 8.}
\end{figure}

Precise measurements of CN, CH$^+$, and CH column densities in the Pleiades 
interstellar gas can offer new insight into the chemical reaction networks 
active in these diffuse clouds. At the same time, highly sensitive 
observations of \ion{Ca}{2} and, to a lesser extent, \ion{Ca}{1} can be used 
to trace the kinematic effects of the interaction occurring between 
interstellar gas and the stellar radiation field in great detail and on large 
spatial scales. The earlier investigation by White (1984\emph{a}), comprising 
spectra for 15 Pleiades members, analyzed interstellar absorption from the 
above atomic and molecular species, but the moderate velocity resolutions 
($\sim$ 3$-$8 km s$^{-1}$) allowed detections of only one component per sight 
line. Later studies capable of detecting many line-of-sight components, such 
as the ultra-high resolution survey by Crane et al. (1995) of interstellar 
CH$^+$ and CH and the high-resolution survey by Welty et al. (1996) of 
\ion{Ca}{2}~K, included only a few of the brightest Pleiades members in their 
sample. Most recently, high-resolution observations were made by White et al. 
(2001) of 36 Pleiades stars, both members and nonmembers, in the \ion{Na}{1}~D 
lines and of 12 of these stars in the \ion{Na}{1} ultraviolet doublet. The 
full analysis of these data (White 2003) revealed considerable complexity and 
star-to-star variation in the atomic gas traced by neutral sodium, leading to 
an extensive spatial schematic of cloud-cluster interactions in which two 
clouds were presumed to be interacting with the UV radiation field of the 
Pleiades and also with each other. One of the primary motivations for the 
present investigation was to obtain equally high quality data for the other 
important optical tracers of the ISM toward a large number of targets in the 
Pleiades so that a complete picture of the interaction between interstellar 
gas clouds and the stars of the cluster may be constructed.

In this paper, high-resolution, high signal to noise observations of CN, 
\ion{Ca}{2}, \ion{Ca}{1}, CH$^+$, and CH toward 20 Pleiades members allow us 
to study cloud-cluster interactions and the implications for the chemistry of 
the local ISM with a precision unmatched by previous investigations. We 
describe our observations and detail the process of data reduction in \S{} 2. 
In \S{} 3, we present the results of Gaussian fitting the observed profiles 
and compare our measurements to those from the literature. The analysis of our 
observations appears in \S{} 4, with particular attention paid to the spatial 
distribution of atomic and molecular velocity components (\S{} 4.1). We 
examine \ion{Na}{1}/\ion{Ca}{2} column density ratios in \S{} 4.2 and derive 
physical conditions in \S{} 4.3. In \S{} 5, we interpret our findings in light 
of the conceptual schematic of cloud-cluster interactions offered by White 
(2003) and summarize our principal results in \S{} 6.

\begin{figure*}
\centering
\includegraphics[width=0.7\textwidth, angle=90]{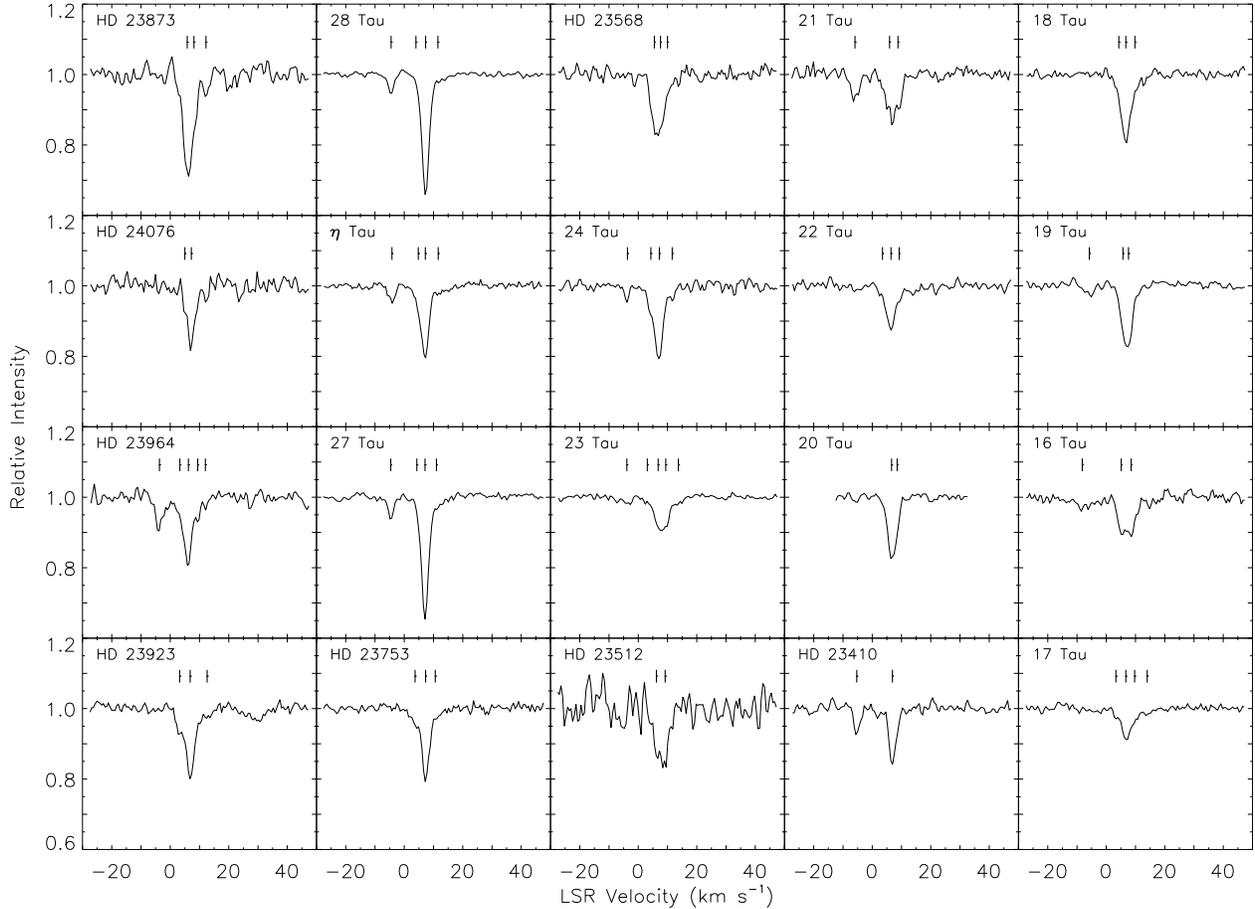}
\caption[Spectra of Ca~{\scriptsize II}.]{Observed Ca~{\scriptsize II} spectra 
arranged to 
approximate the spatial relationships among sight lines. Spectra toward the 
left (right) of the figure correspond to sight lines in the eastern (western) 
part of the cluster. Likewise, spectra toward the top (bottom) of the figure 
correspond to sight lines in the north (south). See Figure 1 for the actual 
distribution of sight lines. The vertical tick marks indicate velocity 
components.}
\end{figure*}

\section{OBSERVATIONS AND DATA REDUCTION}

We observed 20 stars in the Pleiades using the high-resolution mode (cs21) of 
the 2dcoud\'e spectrograph (Tull et al. 1995) on the Harlan J. Smith 2.7 m 
telescope at McDonald Observatory in January, 2002. The choice of sight lines 
came from the list in White et al. (2001), from which only those stars with 
secure membership in the Pleiades cluster were included in our survey. Stellar 
and observational data for the stars in our sample appear in Table 1. 
Specifically, the HD number, name, spectral type, reddening, $B$ magnitude, 
equatorial coordinates (J2000), distance, and total exposure time on each star 
are presented. The spectral types and $E$($B-V$) values come from White et al. 
(2001), while the $B$ magnitudes and equatorial coordinates were obtained from 
the SIMBAD database, operated at CDS, Strasbourg, France. It is most 
appropriate to list $B$ magnitudes because our instrumental setup had a 
central wavelength near 4000 \AA. The parallax for each star was obtained from 
\emph{Hipparcos} measurements (Perryman 1997), when available, and used to 
calculate the distance given in Table 1. Additional information on the program 
stars is available from Table 2 of White et al. (2001). For reference, 
Figure 1 shows the distribution of observed sight lines in the Pleiades 
centered on $\eta$ Tau.

The cross-dispersed echelle spectrometer of the 2dcoud\'e spectrograph was 
configured in a manner identical to that described by Pan et al. (2004) for 
their McDonald Observatory spectra. When combined with a 2048 $\times$ 2048 
CCD, this configuration made possible the search for interstellar absorption 
from many species with a single exposure. In particular, the spectral coverage 
allowed for the detection of absorption features from CN near 3874 \AA, 
\ion{Ca}{2}~K at 3933 \AA, \ion{Ca}{1} $\lambda$4226, CH$^{+}$ $\lambda$4232, 
and CH $\lambda$4300. Each star was observed at least twice with a maximum 
exposure time of 30$^m$ per frame to minimize the effect of cosmic rays. 
Calibration exposures for dark current were obtained the first night of the 
run, while exposures for bias-correction and flat-fielding were taken each 
night and Th-Ar comparison spectra were obtained every 2 to 3 hours. By 
measuring the width of thorium emission lines in the comparison spectra, the 
resolution for these observations is determined to be about 1.7 km s$^{-1}$, 
equivalent to a resolving power of $R\sim175,000$.

\begin{figure*}
\centering
\includegraphics[width=0.7\textwidth, angle=90]{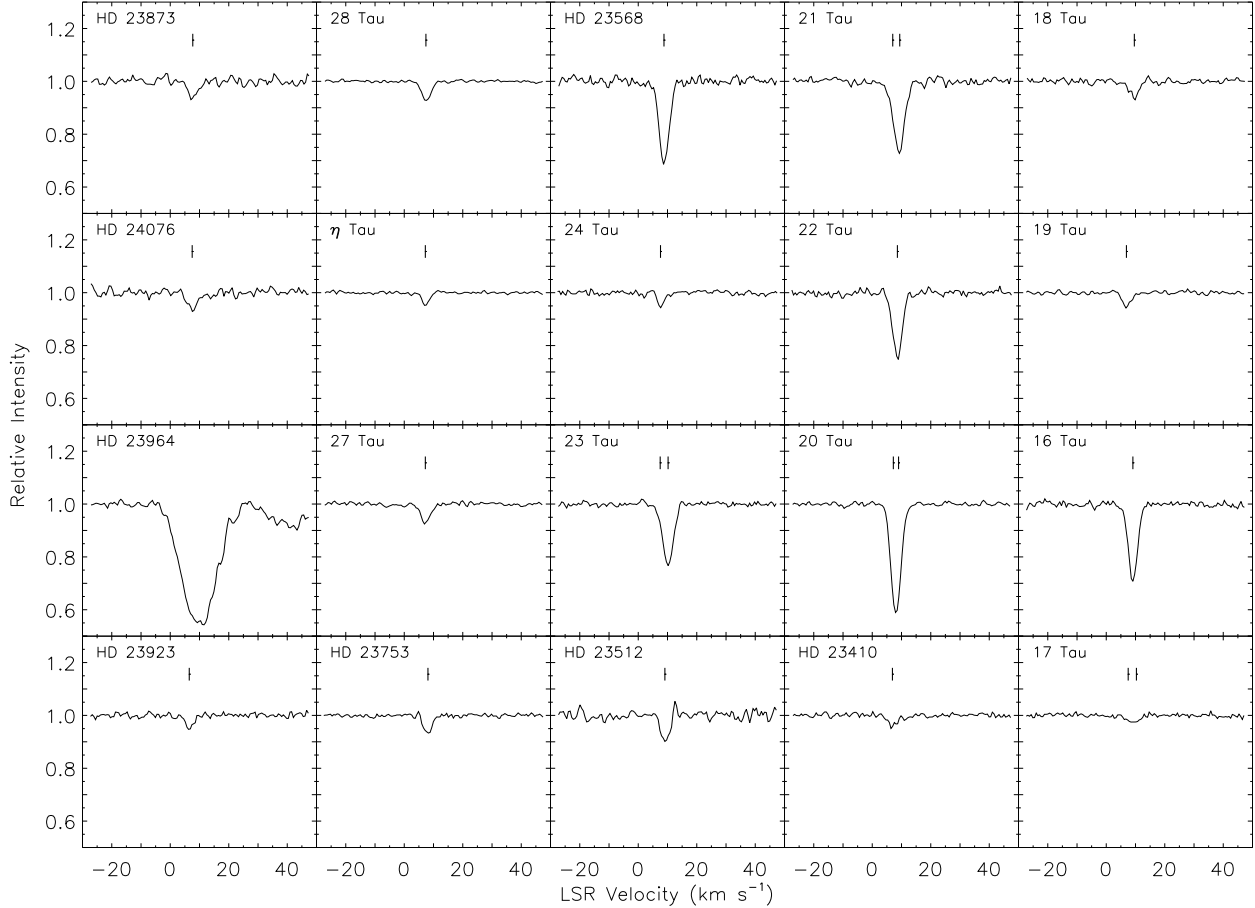}
\caption[Spectra of CH$^{+}$.]{Observed CH$^{+}$ spectra arranged as in Fig. 
2. Strong stellar features in the spectrum of HD 23964 prevented us from 
discerning any interstellar lines here and in Fig. 4.}
\end{figure*}

Standard IRAF routines were used for bias-correction, cosmic-ray removal, and 
flat-fielding of the raw image data. One-dimensional spectra were extracted 
from the processed images and were calibrated in wavelength using the Th-Ar 
comparison spectra and then Doppler-corrected. In regions where interstellar 
absorption lines were expected, individual spectra were cut from each stellar 
exposure with sufficient continua on both sides, typically 2 \AA. All spectra 
for the same species toward the same star were summed to yield a higher 
signal-to-noise ratio (SNR) in the final spectrum. The SNRs obtained in this 
way yielded 3-$\sigma$ upper limits of about 2 m\AA\ on CN absorption and 
about 1 m\AA\ above 4000 \AA. Further details concerning the many steps in the 
data reduction process can be found by consulting Pan et al. (2004), with the 
caveat that no smoothing or additional processing was performed on the data 
presented here. Stellar spectra were normalized to unity by fitting Legendre 
polynomials, of the lowest possible order, to regions free of interstellar 
absorption. This process proceded without complication except in the case of 
HD 23964. The spectra for all of our targeted species toward this star showed 
strong stellar features near the position of the expected interstellar lines. 
For \ion{Ca}{2}~K, we were able to normalize the stellar feature because of 
the strength of the interstellar lines and their location away from the core 
of the stellar profile. But for CN, \ion{Ca}{1}, CH$^{+}$, and CH, the stellar 
features could not be removed by normalization.

Figures 2$-$4 present the final normalized spectra for \ion{Ca}{2}~K, 
CH$^{+}$, and CH, respectively. The panels are arranged to approximate the 
spatial relationship among sight lines so that a comparison might be made of 
absorption profiles in different regions of the cluster (see Fig. 2). Figure 
5 presents the final normalized spectrum for CN absorption toward HD 23512, 
the only sight line in our sample with a detectable amount of this molecule. 
Because, no \ion{Ca}{1} absorption (stronger than $\sim$ 1 m\AA{}) is detected 
along any of the observed sight lines, the spectra for this species are not 
given. These negative results are not unexpected considering the survey of 
\ion{Ca}{1} by Welty, Hobbs, \& Morton (2003) which found predominantly quite 
weak \ion{Ca}{1} lines.

\section{OBSERVATIONAL RESULTS}

The resolution and signal to noise of our observations were high enough to 
allow the detection of multiple components of \ion{Ca}{2}~K and CH$^+$ and to 
permit the first positive detections of CN and CH toward a few of the observed 
stars. The lines were relatively weak, enabling us to fit simple Gaussians to 
the components, yielding values for the equivalent width ($W_{\lambda}$), 
radial velocity in the local standard of rest 
($v_{\mathrm{\scriptstyle LSR}}$), and full width at half-maximum (FWHM) for 
each component. The Doppler parameter ($b$-value) was obtained from the 
measured line width (FWHM), corrected for the instrumental width determined 
from thorium emission lines in the Th-Ar spectra. Uncertainties in 
$W_{\lambda}$ were derived from the rms variations in the normalized continuum 
and the width of the line for a specific component. They do not include errors 
in the placement of the continuum, but this is assumed to have a negligible 
effect. When no component was detected, 3-$\sigma$ upper limits were 
calculated by adopting a typical width for the species in question. Subsequent 
profile synthesis including $\Lambda$-doubling for the CH features yielded no 
significant change in fitted $W_{\lambda}$. Thus, the analysis of this paper 
rests solely on the results of our Gaussian fits.

\begin{figure*}
\centering
\includegraphics[width=0.7\textwidth, angle=90]{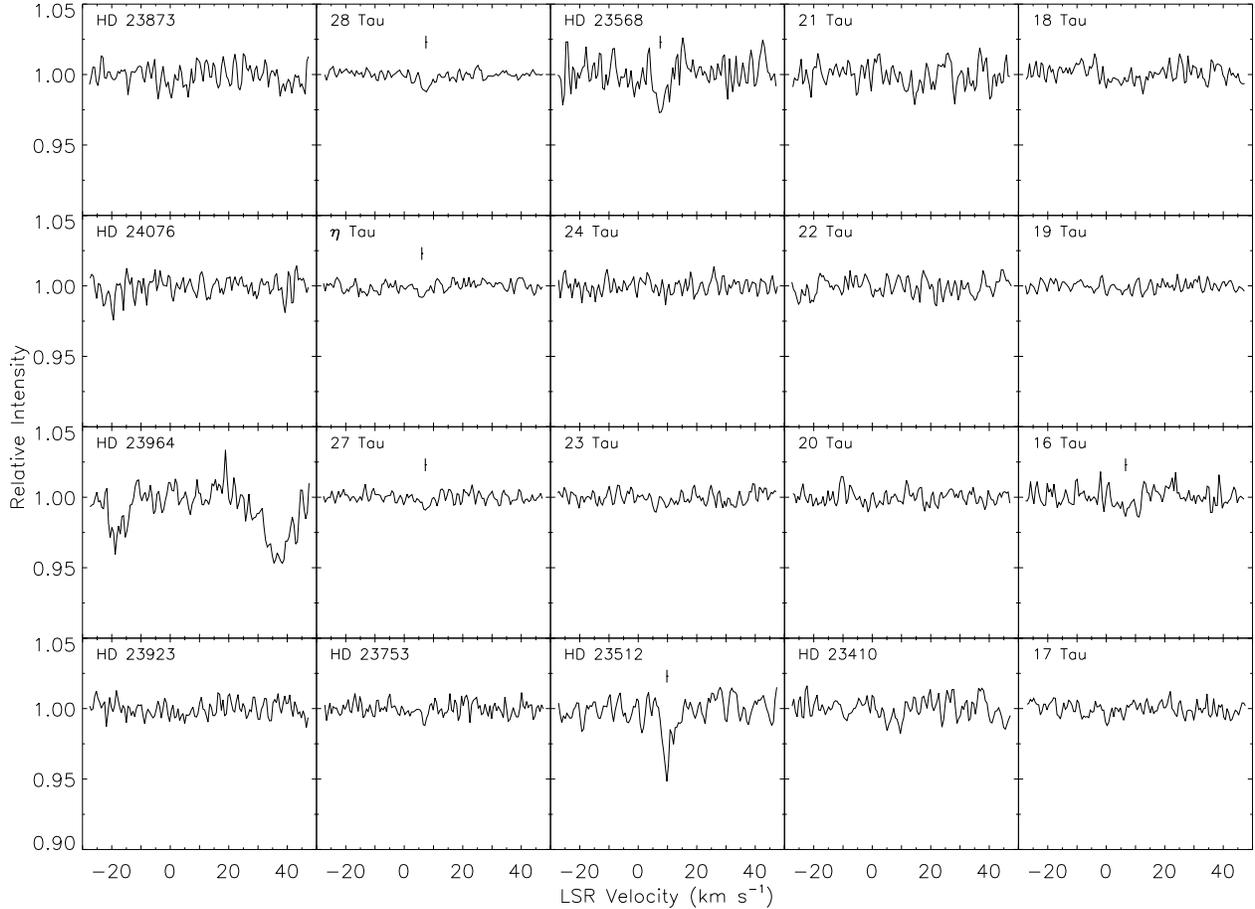}
\caption[Spectra of CH.]{Observed CH spectra arranged as in Fig. 2. For HD 
23512, the intensity scale has been reduced by a factor of 2.}
\end{figure*}

The moderately complex component structures of the \ion{Ca}{2}~K and CH$^+$ 
absorption profiles were constrained by the derived $b$-values. In general, no 
component narrower than 0.4 km s$^{-1}$ or broader than 2.5 km s$^{-1}$ was 
accepted, where the upper limit to $b$ is 1.5 times the mean value for 
\ion{Ca}{2} found by Pan et al. (2005). We note that this restriction plays at 
most a minor role in the component analysis. In some cases, it was necessary 
to compare the velocity of a CH$^+$ component to that of \ion{Ca}{2}~K to 
determine its legitimacy. As discussed in some detail in Pan et al. (2004), 
because \ion{Ca}{2} is the most widely distributed of the observed species, if 
any other species has a component at a particular $v_{\mathrm{\scriptstyle 
LSR}}$, then \ion{Ca}{2} should also have a component at that velocity. Based 
on our resolution, an agreement within 1.5 km s$^{-1}$ was deemed adequate. 
The final component structure for a given line of sight resulted from using 
the fewest number of components, subject to the above criteria, which 
adequately fit the observed profile and left the residuals indistinguishable 
from the noise in the continuum. Naturally, this approach presumes that the 
velocity structure along a line of sight is constant across species. To test 
this assumption, we computed the difference between the velocities found for 
associated CH$^+$ and \ion{Ca}{2} components, $v$(CH$^+)-v$(\ion{Ca}{2}). The 
average difference in velocity for components detected in both species is +0.3 
km s$^{-1}$. To the extent that this result is comparable to the estimated 
uncertainty ($\sigma_v\sim$ 0.2$-$0.3 km s$^{-1}$), we can be reasonably 
assured that our fits reveal the true component structure for a given line of 
sight.

\begin{figure}
\centering
\includegraphics[width=0.4\textwidth]{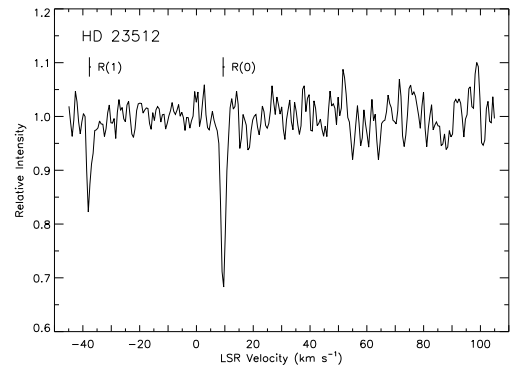}
\caption[Spectrum of CN toward HD 23512.]{Observed CN spectrum toward HD 
23512. The zero point in velocity corresponds to the central wavelength of 
the R(0) line. The zero points of the R(1) and P(1) lines would occur at 
$v_{\mathrm{\scriptstyle LSR}} = -47.2$ km s$^{-1}$ and 
$v_{\mathrm{\scriptstyle LSR}} = +89.0$ km s$^{-1}$ on this scale, 
respectively. The figure shows positive detections of both the R(0) and R(1) 
lines. The P(1) line is below the detection limit.}
\end{figure}

\tabletypesize{\scriptsize}
\begin{deluxetable}{cccccc}
\tablecolumns{6}
\tablewidth{0.4\textwidth}
\tablenum{2}
\tablecaption{Comparison of Total Equivalent Widths}
\tablehead{ \colhead{Ref.} & \colhead{$W_{\lambda}$(Ca~{\tiny II})} & 
\colhead{$W_{\lambda}$(Ca~{\tiny I})} & \colhead{$W_{\lambda}$(CH$^{+}$)} & 
\colhead{$W_{\lambda}$(CH)} & \colhead{$W_{\lambda}$(CN)} \\
\colhead{} & \colhead{(m\AA)} & \colhead{(m\AA)} & \colhead{(m\AA)} & 
\colhead{(m\AA)} & \colhead{(m\AA)} }
\startdata
\multicolumn{6}{c}{16 Tau} \\
\hline
1 & 10.7 $\pm$ 0.7 & $\leq$ 0.9 & 16.0 $\pm$ 0.4 &  0.5 $\pm$ 0.3 & $\leq$ 
1.8 \\
2 & 10.8 $\pm$ 1.4 & $\leq$ 2.4 & 19.5 $\pm$ 1.1 & $\leq$ 4.4 & $\leq$ 5.5 \\
3 & 15 $\pm$ 3 & \ldots & 20 $\pm$ 2 & $\leq$ 13 & \ldots \\
4 & \ldots & \ldots & 18.5 & \ldots & \ldots \\
\hline
\multicolumn{6}{c}{17 Tau} \\
\hline
1 & 6.5 $\pm$ 0.5 & $\leq$ 0.7 & 1.9 $\pm$ 0.3 & $\leq$ 0.6 & $\leq$ 1.4 \\
2 & 6.0 $\pm$ 1.5 & $\leq$ 3.5 & $\leq$ 2.4 & $\leq$ 2.0 & $\leq$ 3.6 \\
5 & 4.4 $\pm$ 0.4 & \ldots & \ldots & \ldots & \ldots \\
6 & 4 & \ldots & \ldots & \ldots & \ldots \\
\hline
\multicolumn{6}{c}{18 Tau} \\
\hline
1 & 11.5 $\pm$ 0.5 & $\leq$ 0.8 & 3.0 $\pm$ 0.3 & $\leq$ 0.8 & $\leq$ 1.2 \\
2 & $\leq$ 13 & $\leq$ 4.0 & $\leq$ 4.6 & $\leq$ 4.0 & $\leq$ 7.7 \\
\hline
\multicolumn{6}{c}{19 Tau} \\
\hline
1 & 11.6 $\pm$ 0.5 & $\leq$ 0.6 & 2.8 $\pm$ 0.2 & $\leq$ 0.5 & $\leq$ 1.2 \\
2 & 12.3 $\pm$ 1.5 & $\leq$ 4.5 & 4.0 $\pm$ 1.3 & $\leq$ 4.1 & $\leq$ 3.7 \\
3 & $\leq$ 13 & \ldots & $\leq$ 13 & $\leq$ 13 & \ldots \\
6 & 9 & \ldots & \ldots & \ldots & \ldots \\
\hline
\multicolumn{6}{c}{20 Tau} \\
\hline
1 & 9.0 $\pm$ 0.4 & $\leq$ 0.8 & 24.4 $\pm$ 0.3 & $\leq$ 0.7 & $\leq$ 1.5 \\
2 & 12.4 $\pm$ 1.1 & $\leq$ 2.2 & 22.2 $\pm$ 0.5 & $\leq$ 1.9 & $\leq$ 2.1 \\
4 & \ldots & \ldots & 20.4 & \ldots & \ldots \\
6 & 8 & \ldots & \ldots & \ldots & \ldots \\
7 & \ldots & \ldots & 24.2 & $\leq$ 0.2 & \ldots \\
8 & \ldots & \ldots & 23.3 $\pm$ 0.3 & \ldots & \ldots \\
9 & \ldots & \ldots & 20.2 $\pm$ 1.8 & 1.3 $\pm$ 0.5 & \ldots \\
10 & \ldots & \ldots & 23.5 & \ldots & \ldots \\
11 & \ldots & \ldots & 31.0 & \ldots & \ldots \\
12 & \ldots & \ldots & 25.4 & \ldots & \ldots \\
\hline
\multicolumn{6}{c}{21 Tau} \\
\hline
1 & 12.4 $\pm$ 0.9 & $\leq$ 1.2 & 18.2 $\pm$ 0.7 & $\leq$ 1.1 & $\leq$ 2.4 \\
2 & 13.4 $\pm$ 1.8 & $\leq$ 6.9 & 20.0 $\pm$ 1.5 & $\leq$ 7.9 & $\leq$ 5.7 \\
3 & 15 $\pm$ 2 & \ldots & 25 $\pm$ 3 & $\leq$ 13 & \ldots \\
\hline
\multicolumn{6}{c}{22 Tau} \\
\hline
1 & 7.3 $\pm$ 0.6 & $\leq$ 1.0 & 14.0 $\pm$ 0.4 & $\leq$ 0.9 & $\leq$ 1.6 \\
2 & 17.7 $\pm$ 3.2 & $\leq$ 8.1 & $\leq$ 8.7 & $\leq$ 6.7 & $\leq$ 10 \\
\hline
\multicolumn{6}{c}{23 Tau} \\
\hline
1 & 9.0 $\pm$ 0.4 & $\leq$ 0.6 & 15.0 $\pm$ 0.4 & $\leq$ 0.6 & $\leq$ 1.3 \\
2 & 10.3 $\pm$ 1.1 & $\leq$ 2.9 & 14.1 $\pm$ 0.5 & $\leq$ 2.6 & $\leq$ 3.5 \\
3 & $\leq$ 13 & \ldots & 25 $\pm$ 5 & $\leq$ 13 & \ldots \\
4 & \ldots & \ldots & 14.2 & \ldots & \ldots \\
6 & 7 & \ldots & \ldots & \ldots & \ldots \\
7 & \ldots & \ldots & 16.2 & $\leq$ 3.1 & \ldots \\
8 & \ldots & \ldots & 14.0 $\pm$ 0.2 & \ldots & \ldots \\
9 & \ldots & \ldots & 12.5 $\pm$ 2.2 & $\leq$ 2.9 & \ldots \\
11 & \ldots & \ldots & 16.0 & \ldots & \ldots \\
\hline
\multicolumn{6}{c}{HD 23512} \\
\hline
1 & 11.2 $\pm$ 1.7 & $\leq$ 2.2 & 5.1 $\pm$ 0.6 & 4.3 $\pm$ 0.6 & 8.7 $\pm$ 
0.7 \\
2 & $\leq$ 22 & $\leq$ 13 & $\leq$ 14 & 23 $\pm$ 5 & $\leq$ 29 \\
13 & \ldots & \ldots & \ldots & 5.3 $\pm$ 0.8 & $\leq$ 10.0 \\
\hline
\multicolumn{6}{c}{HD 23568} \\
\hline
1 & 12.5 $\pm$ 1.0 & $\leq$ 1.6 & 18.4 $\pm$ 0.6 & 1.3 $\pm$ 0.4 & $\leq$ 3.3 
\\
2 & 15.6 $\pm$ 3.0 & $\leq$ 7.9 & 15.5 $\pm$ 2.6 & $\leq$ 7.7 & $\leq$ 8.7 \\
\hline
\multicolumn{6}{c}{$\eta$ Tau} \\
\hline
1 & 12.5 $\pm$ 0.4 & $\leq$ 0.4 & 2.0 $\pm$ 0.1 & 0.4 $\pm$ 0.1 & $\leq$ 0.9 \\
2 & 15.0 $\pm$ 1.1 & $\leq$ 3.3 & 3.8 $\pm$ 0.7 & $\leq$ 2.8 & $\leq$ 3.7 \\
5 & 12.1 $\pm$ 0.7 & \ldots & \ldots & \ldots & \ldots \\
5 & 14.4 $\pm$ 1.2 & \ldots & \ldots & \ldots & \ldots \\
6 & 11 & \ldots & \ldots & \ldots & \ldots \\
7 & \ldots & \ldots & 1.7 & \ldots & \ldots \\
9 & \ldots & \ldots & 1.6 $\pm$ 0.4 & $\leq$ 1.8 & \ldots \\
14 & 13.6 $\pm$ 0.2 & \ldots & \ldots & \ldots & \ldots \\
15 & 8 & $\leq$ 0.64 & \ldots & \ldots & \ldots \\
\hline
\multicolumn{6}{c}{HD 23753} \\
\hline
1 & 11.0 $\pm$ 0.6 & $\leq$ 0.9 & 3.5 $\pm$ 0.2 & $\leq$ 0.7 & $\leq$ 1.2 \\
2 & 10.6 $\pm$ 1.0 & $\leq$ 4.2 & 4.0 $\pm$ 0.9 & $\leq$ 4.5 & $\leq$ 5.5 \\
4 & \ldots & \ldots & 7.6 & \ldots & \ldots \\
\hline
\multicolumn{6}{c}{27 Tau} \\
\hline
1 & 17.7 $\pm$ 0.4 & $\leq$ 0.7 & 4.1 $\pm$ 0.3 & 0.4 $\pm$ 0.2 & $\leq$ 1.1 \\
2 & 12.1 $\pm$ 1.2 & $\leq$ 3.9 & 5.0 $\pm$ 0.7 & $\leq$ 3.1 & $\leq$ 4.1 \\
3 & 13 $\pm$ 1 & \ldots & $\leq$ 13 & $\leq$ 13 & \ldots \\
6 & 12 & \ldots & \ldots & \ldots & \ldots \\
7 & \ldots & \ldots & 4.2 & \ldots & \ldots \\
\hline
\multicolumn{6}{c}{28 Tau} \\
\hline
1 & 17.0 $\pm$ 0.3 & $\leq$ 0.5 & 4.1 $\pm$ 0.2 & 0.6 $\pm$ 0.1 & $\leq$ 0.8 \\
2 & 13.8 $\pm$ 1.6 & $\leq$ 4.5 & $\leq$ 2.9 & $\leq$ 3.5 & $\leq$ 5.1 \\
\enddata
\end{deluxetable}
\tabletypesize{\scriptsize}
\begin{deluxetable}{cccccc}
\tablecolumns{6}
\tablewidth{0.4\textwidth}
\tablenum{2}
\tablecaption{---\emph{Continued}}
\tablehead{ \colhead{Ref.} & \colhead{$W_{\lambda}$(Ca~{\tiny II})} & 
\colhead{$W_{\lambda}$(Ca~{\tiny I})} & \colhead{$W_{\lambda}$(CH$^{+}$)} & 
\colhead{$W_{\lambda}$(CH)} & \colhead{$W_{\lambda}$(CN)} \\
\colhead{} & \colhead{(m\AA)} & \colhead{(m\AA)} & \colhead{(m\AA)} & 
\colhead{(m\AA)} & \colhead{(m\AA)} }
\startdata
\multicolumn{6}{c}{HD 23923} \\
\hline
1 & 12.7 $\pm$ 0.7 & $\leq$ 1.0 & 2.4 $\pm$ 0.3 & $\leq$ 0.8 & $\leq$ 1.4 \\
2 & 16.9 $\pm$ 1.5 & $\leq$ 5.8 & $\leq$ 11 & $\leq$ 5.1 & $\leq$ 11 \\
\enddata
\tablerefs{(1) Present investigation; (2) White 1984\emph{a}; (3) Younan \& 
Dufton 1984; (4) Frisch 1972; (5) Welty et al. 1996; (6) Marschall \& Hobbs 
1972; (7) Crane et al. 1995; (8) Hawkins \& Jura 1987; (9) Federman 1982; (10) 
Vanden Bout \& Snell 1980; (11) Hobbs 1973; (12) Vanden Bout \& Thaddeus 1971; 
(13) Federman et al. 1994; (14) Vallerga et al. 1993; (15) White 1973.}
\end{deluxetable}

As seen in Figure 2, interstellar absorption from \ion{Ca}{2}~K is detected 
along all twenty sight lines, resulting in 65 individual components. We note 
that our study detects 50 percent more absorption components in \ion{Ca}{2} 
than White et al. (2001) found in their \ion{Na}{1} observations for the same 
directions. Figure 3 shows that CH$^+$ absorption is just as pervasive, yet 
only 23 components are identified. No detection of CH$^+$ toward HD 23964 is 
possible because the core of the stellar line discussed in \S{} 2 directly 
coincides with the expected position of the interstellar feature. Marginal 
detections of CH toward five of the observed stars are indicated in Figure 4. 
A stronger component is detected toward HD 23512, a sight line which passes 
through a known molecular cloud (Federman \& Willson 1984). The spectrum of CN 
toward this star (Figure 5) clearly shows the absorption features of the R(0) 
and R(1) transitions of this molecule, but the weaker P(1) line is below the 
detection limit.

\tabletypesize{\scriptsize}
\begin{deluxetable}{lccc}
\tablecolumns{4}
\tablewidth{0.4\textwidth}
\tablenum{3}
\tablecaption{Wavelengths (Air) and $f$-values}
\tablehead{ \colhead{Species} & \colhead{$\lambda_{\mathrm{air}}$} & 
\colhead{$f$-value} & \colhead{Ref.} \\
\colhead{} & \colhead{(\AA)} & \colhead{} & \colhead{} }
\startdata
Ca~{\tiny II} & 3933.663 & 0.6346\phn & 1 \\
CH$^{+}$ & 4232.548 & 0.00545 & 2 \\
CH & 4300.313 & 0.00510 & 3 \\
CN\tablenotemark{a} & 3874.00\phn & 0.0228\phn & 3 \\
 & 3874.61\phn & 0.0342\phn & 3 \\
 & 3875.76\phn & 0.0114\phn & 3 \\
\enddata
\tablenotetext{a}{First entry is for the R(1) transition, second is for R(0), 
third is for P(1).}
\tablerefs{(1) Morton 1991; (2) Gredel et al. 1993; (3) Federman et al. 1994.}
\end{deluxetable}

\tabletypesize{\scriptsize}
\begin{deluxetable*}{cccccccccccccccc}
\tablecolumns{16}
\tablewidth{\textwidth}
\tablenum{4}
\setlength{\tabcolsep}{1pt}
\tablecaption{Summary of Results for Individual Components}
\tablehead{\multicolumn{3}{c}{Ca~{\tiny II}} & \colhead{} & 
\multicolumn{3}{c}{CH$^{+}$} & \colhead{} & \multicolumn{3}{c}{CH} & 
\colhead{} & \multicolumn{3}{c}{Na~{\tiny I}\tablenotemark{a}} & 
\colhead{} \\
\cline{1-3} \cline{5-7} \cline{9-11} \cline{13-16} \\
\colhead{$N$} & \colhead{$v_{\mathrm{{\scriptscriptstyle LSR}}}$} & 
\colhead{$b$-value} & 
\colhead{} & \colhead{$N$} & \colhead{$v_{\mathrm{{\scriptscriptstyle LSR}}}$} 
& \colhead{$b$-value} & \colhead{} & \colhead{$N$} & 
\colhead{$v_{\mathrm{{\scriptscriptstyle LSR}}}$} & \colhead{$b$-value} & 
\colhead{} & \colhead{$N$} & \colhead{$v_{\mathrm{{\scriptscriptstyle LSR}}}$} 
& \colhead{$b$-value} & \colhead{$N$(Na~{\tiny I})/$N$(Ca~{\tiny II})} \\
\colhead{($10^{10}$ cm$^{-2}$)} & \colhead{(km s$^{-1}$)} & 
\colhead{(km s$^{-1}$)} & \colhead{} & \colhead{($10^{12}$ cm$^{-2}$)} & 
\colhead{(km s$^{-1}$)} & \colhead{(km s$^{-1}$)} & \colhead{} & 
\colhead{($10^{12}$ cm$^{-2}$)} & \colhead{(km s$^{-1}$)} & 
\colhead{(km s$^{-1}$)} & \colhead{} & \colhead{($10^{11}$ cm$^{-2}$)} & 
\colhead{(km s$^{-1}$)} & \colhead{(km s$^{-1}$)} & \colhead{} }
\startdata
\multicolumn{16}{c}{16 Tau} \\
\hline
1.8 $\pm$ 0.5 & $-8.2$ & 1.9 && $\leq$ 1.8 & \ldots & \ldots && $\leq$ 1.2 & 
\ldots & \ldots && $\leq$ 0.9 & \ldots & \ldots & $\leq$ 5.0 \\
4.6 $\pm$ 0.5 & +5.1 & 1.4 && $\leq$ 1.8 & \ldots & \ldots && 0.6 $\pm$ 0.4 & 
+6.6 & 1.6 && 2.8 $\pm$ 0.3 & +6.4 & 1.3 & 6.1 $\pm$ 0.9 \\
6.4 $\pm$ 0.5 & +8.5 & 2.0 && 21.7 $\pm$ 0.6 & +9.1 & 1.9 && $\leq$ 1.2 & 
\ldots & \ldots && 2.5 $\pm$ 0.3 & +9.3 & 1.1 & 3.9 $\pm$ 0.6 \\
\hline
\multicolumn{16}{c}{17 Tau} \\
\hline
0.8 $\pm$ 0.2 & +3.3 & 0.4 && $\leq$ 0.6 & \ldots & \ldots && $\leq$ 0.7 & 
\ldots & \ldots && $\leq$ 0.2 & \ldots & \ldots & $\leq$ 2.5 \\
4.3 $\pm$ 0.2 & +6.7 & 1.4 && 0.9 $\pm$ 0.2 & +7.5 & 1.6 && $\leq$ 0.7 & 
\ldots & \ldots && 2.0 $\pm$ 0.1 & +7.0 & 1.1 & 4.6 $\pm$ 0.3 \\
1.8 $\pm$ 0.4 & +9.7 & 1.5 && 1.3 $\pm$ 0.2 & +10.3 & 1.6 && $\leq$ 0.7 & 
\ldots & \ldots && 0.4 $\pm$ 0.1 & +9.4 & 0.7 & 2.2$\pm$ 0.7 \\
0.8 $\pm$ 0.2 & +14.0 & 1.3 && $\leq$ 0.6 & \ldots & \ldots && $\leq$ 0.7 & 
\ldots & \ldots && $\leq$ 0.2 & \ldots & \ldots & $\leq$ 2.5 \\
\hline
\multicolumn{16}{c}{18 Tau} \\
\hline
1.9 $\pm$ 0.4 & +4.3 & 1.6 && $\leq$ 1.2 & \ldots & \ldots && $\leq$ 1.0 & 
\ldots & \ldots && $\leq$ 0.4 & \ldots & \ldots & $\leq$ 2.1 \\
9.8 $\pm$ 0.4 & +6.7 & 1.6 && $\leq$ 1.2 & \ldots & \ldots && $\leq$ 1.0 & 
\ldots & \ldots && 3.6 $\pm$ 0.5 & +7.4 & 1.0 & 3.7 $\pm$ 0.5 \\
2.3 $\pm$ 0.4 & +9.8 & 1.6 && 3.6 $\pm$ 0.4 & +9.6 & 1.5 && $\leq$ 1.0 & 
\ldots & \ldots && 0.8 $\pm$ 0.1 & +10.9 & 0.9 & 3.5 $\pm$ 0.7 \\
\hline
\multicolumn{16}{c}{19 Tau} \\
\hline
1.5 $\pm$ 0.4 & $-5.8$ & 1.8 && $\leq$ 0.6 & \ldots & \ldots && $\leq$ 0.6 & 
\ldots & \ldots && $\leq$ 3.0 & \ldots & \ldots & $\leq$ 20.0 \\
3.4 $\pm$ 0.4 & +5.7 & 1.6 && 3.3 $\pm$ 0.2 & +6.9 & 1.8 && $\leq$ 0.6 & 
\ldots & \ldots && 4.0 $\pm$ 1.0 & +6.7 & 1.0 & 11.8 $\pm$ 3.2 \\
9.2 $\pm$ 0.4 & +7.6 & 1.9 && $\leq$ 0.6 & \ldots & \ldots && $\leq$ 0.6 & 
\ldots & \ldots && 6.0 $\pm$ 1.1 & +8.3 & 0.8 & 6.5 $\pm$ 1.2 \\
\hline
\multicolumn{16}{c}{20 Tau} \\
\hline
9.3 $\pm$ 0.4 & +6.6 & 1.7 && 15.7 $\pm$ 0.3 & +7.2 & 1.5 && $\leq$ 0.8 & 
\ldots & \ldots && 8.0 $\pm$ 0.4 & +6.9 & 1.0 & 8.6 $\pm$ 0.6 \\
1.8 $\pm$ 0.2 & +8.5 & 0.9 && 16.0 $\pm$ 0.3 & +9.0 & 1.6 && $\leq$ 0.8 & 
\ldots & \ldots && 2.4 $\pm$ 0.2 & +8.9 & 0.9 & 13.3 $\pm$ 1.8 \\
\hline
\multicolumn{16}{c}{HD 23410} \\
\hline
2.8 $\pm$ 0.4 & $-5.3$ & 0.9 && $\leq$ 1.2 & \ldots & \ldots && $\leq$ 1.2 & 
\ldots & \ldots && 0.6 $\pm$ 0.2 & $-$5.2 & 0.2 & 2.1 $\pm$ 0.8 \\
7.7 $\pm$ 0.5 & +6.9 & 1.4 && 2.7 $\pm$ 0.4 & +6.9 & 2.1&& $\leq$ 1.2 & \ldots 
& \ldots && 12. $\pm$ 0.8 & +7.1 & 1.1 & 15.6 $\pm$ 1.4 \\
\hline
\multicolumn{16}{c}{21 Tau} \\
\hline
3.8 $\pm$ 0.6 & $-5.9$ & 1.6 && $\leq$ 2.1 & \ldots & \ldots && $\leq$ 1.3 & 
\ldots & \ldots && $\leq$ 1.9 & \ldots & \ldots & $\leq$ 5.0 \\
6.8 $\pm$ 0.8 & +5.9 & 2.2 && 4.9 $\pm$ 0.6 & +7.0 & 2.3 && $\leq$ 1.3 & 
\ldots & \ldots && 4.1 $\pm$ 0.7 & +6.9 & 1.0 & 6.0 $\pm$ 1.2 \\
4.3 $\pm$ 0.6 & +8.8 & 1.8 && 18.7 $\pm$ 0.8 & +9.4 & 2.1 && $\leq$ 1.3 & 
\ldots & \ldots && 1.3 $\pm$ 0.6 & +9.2 & 1.1 & 3.0 $\pm$ 1.4 \\
\hline
\multicolumn{16}{c}{22 Tau} \\
\hline
1.3 $\pm$ 0.5 & +3.5 & 1.4 && $\leq$ 1.8 & \ldots & \ldots && $\leq$ 1.1 & 
\ldots & \ldots && $\leq$ 1.4 & \ldots & \ldots & $\leq$ 10.8 \\
5.9 $\pm$ 0.5 & +6.4 & 1.4 && $\leq$ 1.8 & \ldots & \ldots && $\leq$ 1.1 & 
\ldots & \ldots && 4.0 $\pm$ 0.5 & +6.8 & 1.0 & 6.8 $\pm$ 1.0 \\
1.5 $\pm$ 0.4 & +9.2 & 1.2 && 18.6 $\pm$ 0.6 & +8.6 & 2.0 && $\leq$ 1.1 & 
\ldots & \ldots && 0.9 $\pm$ 0.4 & +9.2 & 1.0 & 6.0 $\pm$ 3.1 \\
\hline
\multicolumn{16}{c}{23 Tau} \\
\hline
1.0 $\pm$ 0.2 & $-3.9$ & 1.9 && $\leq$ 1.2 & \ldots & \ldots && $\leq$ 0.7 & 
\ldots & \ldots && $\leq$ 0.4 & \ldots & \ldots & $\leq$ 4.0 \\
0.7 $\pm$ 0.2 & +3.1 & 1.5 && $\leq$ 1.2 & \ldots & \ldots && $\leq$ 0.7 & 
\ldots & \ldots && $\leq$ 0.4 & \ldots & \ldots & $\leq$ 5.7 \\
4.2 $\pm$ 0.2 & +6.8 & 1.7 && 1.3 $\pm$ 0.4 & +7.5 & 2.3 && $\leq$ 0.7 & 
\ldots & \ldots && 3.0 $\pm$ 0.5 & +7.5 & 1.0, 3.4\tablenotemark{b} & 7.1 
$\pm$ 1.2 \\
3.8 $\pm$ 0.2 & +9.4 & 1.6 && 18.4 $\pm$ 0.4 & +10.2 & 2.2 && $\leq$ 0.7 & 
\ldots & \ldots && 1.0 $\pm$ 0.1 & +10.6 & 0.6 & 2.6 $\pm$ 0.3 \\
0.9 $\pm$ 0.2 & +13.7 & 1.5  && $\leq$ 1.2 & \ldots & \ldots && $\leq$ 0.7 & 
\ldots & \ldots && $\leq$ 0.4 & \ldots & \ldots & $\leq$ 4.4 \\
\hline
\multicolumn{16}{c}{HD 23512} \\
\hline
5.8 $\pm$ 1.5 & +6.2 & 1.5 && $\leq$ 2.4 & \ldots & \ldots && $\leq$ 2.4 & 
\ldots & \ldots && 5.9 $\pm$ 3.1 & +7.4 & 0.5 & 10.2 $\pm$ 6.0 \\
7.8 $\pm$ 1.6 & +9.2 & 1.6 && 6.2 $\pm$ 0.8 & +9.1 & 1.7 && 5.4 $\pm$ 0.8 & 
+9.8 & 1.6 && $>$ 60. & +10.1 & 0.6 & $>$ 76.9 \\
\hline
\multicolumn{16}{c}{HD 23568} \\
\hline
6.3 $\pm$ 0.6 & +5.5 & 1.6 && $\leq$ 3.0 & \ldots & \ldots && $\leq$ 1.5 & 
\ldots & \ldots && $\leq$ 2.5 & \ldots & \ldots & $\leq$ 4.0 \\
6.3 $\pm$ 0.6 & +7.6 & 1.8 && $\leq$ 3.0 & \ldots & \ldots && 1.6 $\pm$ 0.5 & 
+7.5 & 1.6 && 9.4 $\pm$ 1.0 & +7.1 & 1.0 & 14.9 $\pm$ 2.1 \\
2.5 $\pm$ 0.8 & +10.0 & 2.5 && 25.6 $\pm$ 1.0 & +8.8 & 2.1 && $\leq$ 1.5 & 
\ldots & \ldots && 1.8 $\pm$ 0.6 & +9.5 & 0.6 & 7.2 $\pm$ 3.3 \\
\hline
\multicolumn{16}{c}{24 Tau} \\
\hline
1.2 $\pm$ 0.2 & $-3.7$ & \ldots && $\leq$ 0.6 & \ldots & \ldots && $\leq$ 0.8 
& \ldots & \ldots && $\leq$ 2.0 & \ldots & \ldots & $\leq$ 16.7 \\
3.1 $\pm$ 0.5 & +4.3 & 1.4 && $\leq$ 0.6 & \ldots & \ldots && $\leq$ 0.8 & 
\ldots & \ldots && $\leq$ 2.0 & \ldots & \ldots & $\leq$ 6.4 \\
11.1 $\pm$ 0.5 & +7.2 & 1.5 && 2.5 $\pm$ 0.2 & +7.6 & 1.0 && $\leq$ 0.8 & 
\ldots & \ldots && 9.7 $\pm$ 0.9 & +7.1 & 2.6, 0.9\tablenotemark{b} & 8.7 
$\pm$ 0.9 \\
1.5 $\pm$ 0.4 & +11.6 & 1.1 && $\leq$ 0.6 & \ldots & \ldots && $\leq$ 0.8 & 
\ldots & \ldots && $\leq$ 2.0 & \ldots & \ldots & $\leq$ 13.3 \\
\hline
\multicolumn{16}{c}{$\eta$ Tau} \\
\hline
2.0 $\pm$ 0.2 & $-4.2$ & 1.2 && $\leq$ 0.3 & \ldots & \ldots && $\leq$ 0.3 & 
\ldots & \ldots && 0.1 $\pm$ 0.02 & $-$3.4 & 1.9 & 0.5 $\pm$ 0.1 \\
2.3 $\pm$ 0.2 & +4.8 & 1.3  && $\leq$ 0.3 & \ldots & \ldots && $\leq$ 0.3 & 
\ldots & \ldots && 0.4 $\pm$ 0.05 & +5.6 & 3.8 & 1.7 $\pm$ 0.3 \\
9.7 $\pm$ 0.3 & +7.2 & 1.3 && 2.4 $\pm$ 0.1 & +7.2 & 1.4 && 0.5 $\pm$ 0.1 & 
+6.0 & 1.5 && 7.5 $\pm$ 0.2 & +7.5 & 1.1 & 7.7 $\pm$ 0.3 \\
1.2 $\pm$ 0.2 & +11.6 & 1.3 && $\leq$ 0.3 & \ldots & \ldots && $\leq$ 0.3 & 
\ldots & \ldots && $\leq$ 0.1 & \ldots & \ldots & $\leq$ 0.8 \\
\hline
\multicolumn{16}{c}{HD 23753} \\
\hline
2.5 $\pm$ 0.5 & +3.7 & 1.7 && $\leq$ 0.6 & \ldots & \ldots && $\leq$ 0.8 & 
\ldots & \ldots && 1.5 $\pm$ 0.2 & +4.0 & 0.3 & 6.0 $\pm$ 1.4 \\
9.2 $\pm$ 0.4 & +7.3 & 1.2 && 4.2 $\pm$ 0.2 & +8.1 & 1.7 && $\leq$ 0.8 & 
\ldots & \ldots && 7.2 $\pm$ 0.3 & +7.6 & 1.2 & 7.8 $\pm$ 0.5 \\
1.8 $\pm$ 0.5 & +10.6 & 2.2 && $\leq$ 0.6 & \ldots & \ldots && $\leq$ 0.8 & 
\ldots & \ldots && $\leq$ 0.5 & \ldots & \ldots & $\leq$ 2.8 \\
\enddata
\end{deluxetable*}

\tabletypesize{\scriptsize}
\begin{deluxetable*}{cccccccccccccccc}
\tablecolumns{16}
\tablewidth{\textwidth}
\tablenum{4}
\setlength{\tabcolsep}{1pt}
\tablecaption{---\emph{Continued}}
\tablehead{\multicolumn{3}{c}{Ca~{\tiny II}} & \colhead{} & 
\multicolumn{3}{c}{CH$^{+}$} & \colhead{} & \multicolumn{3}{c}{CH} & 
\colhead{} & \multicolumn{3}{c}{Na~{\tiny I}\tablenotemark{a}} & 
\colhead{} \\
\cline{1-3} \cline{5-7} \cline{9-11} \cline{13-16} \\
\colhead{$N$} & \colhead{$v_{\mathrm{{\scriptscriptstyle LSR}}}$} & 
\colhead{$b$-value} & 
\colhead{} & \colhead{$N$} & \colhead{$v_{\mathrm{{\scriptscriptstyle LSR}}}$} 
& \colhead{$b$-value} & \colhead{} & \colhead{$N$} & 
\colhead{$v_{\mathrm{{\scriptscriptstyle LSR}}}$} & \colhead{$b$-value} & 
\colhead{} & \colhead{$N$} & \colhead{$v_{\mathrm{{\scriptscriptstyle LSR}}}$} 
& \colhead{$b$-value} & \colhead{$N$(Na~{\tiny I})/$N$(Ca~{\tiny II})} \\
\colhead{($10^{10}$ cm$^{-2}$)} & \colhead{(km s$^{-1}$)} & 
\colhead{(km s$^{-1}$)} & \colhead{} & \colhead{($10^{12}$ cm$^{-2}$)} & 
\colhead{(km s$^{-1}$)} & \colhead{(km s$^{-1}$)} & \colhead{} & 
\colhead{($10^{12}$ cm$^{-2}$)} & \colhead{(km s$^{-1}$)} & 
\colhead{(km s$^{-1}$)} & \colhead{} & \colhead{($10^{11}$ cm$^{-2}$)} & 
\colhead{(km s$^{-1}$)} & \colhead{(km s$^{-1}$)} & \colhead{} }
\startdata
\multicolumn{16}{c}{27 Tau} \\
\hline
2.2 $\pm$ 0.2 & $-4.6$ & 0.8 && $\leq$ 1.2 & \ldots &\ldots  && $\leq$ 0.6 & 
\ldots & \ldots && 0.1 $\pm$ 0.01 & $-$4.1 & 1.6 & 0.4 $\pm$ 0.1 \\
0.9 $\pm$ 0.2 & +4.3 & 1.4 && $\leq$ 1.2 & \ldots &\ldots  && $\leq$ 0.6 & 
\ldots & \ldots && $\leq$ 0.03 & \ldots & \ldots & $\leq$ 0.3 \\
17.7 $\pm$ 0.3 & +7.1 & 1.3 && 4.9 $\pm$ 0.4 & +7.2 & 2.1 && 0.5 $\pm$ 0.2 & 
+7.2 & 1.6 && 20. $\pm$ 1.1 & +7.2 & 3.8, 0.9\tablenotemark{b} & 11.3 $\pm$ 
0.6 \\
1.9 $\pm$ 0.2 & +11.0 & 1.8  && $\leq$ 1.2 & \ldots & \ldots && $\leq$ 0.6 & 
\ldots & \ldots && 0.7 $\pm$ 0.2 & +9.0 & 0.6 & 3.7 $\pm$ 1.1 \\
\hline
\multicolumn{16}{c}{28 Tau} \\
\hline
2.0 $\pm$ 0.1 & $-4.5$ & 1.0 && $\leq$ 0.6 & \ldots &\ldots  && $\leq$ 0.3 & 
\ldots & \ldots && $\leq$ 0.6 & \ldots & \ldots & $\leq$ 3.0 \\
0.7 $\pm$ 0.1 & +4.0 & 1.2 && $\leq$ 0.6 & \ldots &\ldots  && $\leq$ 0.3 & 
\ldots & \ldots && $\leq$ 0.6 & \ldots & \ldots & $\leq$ 8.6 \\
17.9 $\pm$ 0.3 & +7.3 & 1.4 && 4.9 $\pm$ 0.2 & +7.4 & 2.0 && 0.7 $\pm$ 0.1 & 
+7.4 & 1.6 && 4.7 $\pm$ 0.2 & +6.8 & 1.0 & 2.6 $\pm$ 0.1 \\
1.4 $\pm$ 0.2 & +11.5 & 2.1 && $\leq$ 0.6 & \ldots &\ldots  && $\leq$ 0.3 & 
\ldots & \ldots && $\leq$ 0.6 & \ldots & \ldots & $\leq$ 4.3 \\
\hline
\multicolumn{16}{c}{HD 23873} \\
\hline
16.2 $\pm$ 1.0 & +5.8 & 1.8 && $\leq$ 2.1 & \ldots & \ldots && $\leq$ 1.2 & 
\ldots & \ldots && 7.9 $\pm$ 0.3 & +6.6 & 1.5 & 4.9 $\pm$ 0.4 \\
5.0 $\pm$ 0.7 & +8.1 & 1.6 && 4.4 $\pm$ 0.7 & +7.7 & 2.1 && $\leq$ 1.2 & 
\ldots & \ldots && $\leq$ 1.0 & \ldots & \ldots & $\leq$ 2.0 \\
2.5 $\pm$  0.6 & +12.2 & 1.1 && $\leq$ 2.1 & \ldots & \ldots && $\leq$ 1.2 & 
\ldots & \ldots && $\leq$ 1.0 & \ldots & \ldots & $\leq$ 4.0 \\
\hline
\multicolumn{16}{c}{HD 23923} \\
\hline
2.9 $\pm$ 0.4 & +3.2 & 1.1 && $\leq$ 1.2 & \ldots & \ldots && $\leq$ 1.0 & 
\ldots & \ldots && 3.0 $\pm$ 0.2 & +3.9 & 1.5 & 10.3 $\pm$ 1.6 \\
10.9 $\pm$ 0.4 & +6.8 & 1.6 && 2.8 $\pm$ 0.4 &  +6.5 & 1.5 && $\leq$ 1.0 & 
\ldots & \ldots && 6.8 $\pm$ 0.4 & +7.1 & 0.8 & 6.2 $\pm$ 0.4 \\
1.8 $\pm$ 0.6 & +12.6 & 2.3 && $\leq$ 1.2 & \ldots & \ldots && $\leq$ 1.0 & 
\ldots & \ldots && $\leq$ 1.0 & \ldots & \ldots & $\leq$ 5.6 \\
\hline
\multicolumn{16}{c}{HD 23964} \\
\hline
4.9 $\pm$ 0.7 & $-3.7$ & 1.8 && $\leq$ 1.3 & \ldots & \ldots && $\leq$ 1.5 & 
\ldots & \ldots && 0.3 $\pm$ 0.04 & $-$3.7 & 0.5 & 0.6 $\pm$0.1 \\
2.6 $\pm$ 0.6 & +3.3 & 1.5  && $\leq$ 1.3 & \ldots & \ldots && $\leq$ 1.5 & 
\ldots & \ldots && 3.1 $\pm$ 0.8 & +4.8 & 2.7 & 11.9 $\pm$ 4.1 \\
9.0 $\pm$ 0.5 & +6.2 & 1.3  && $\leq$ 1.3 & \ldots & \ldots && $\leq$ 1.5 & 
\ldots & \ldots && 7.5 $\pm$ 0.8 & +6.7 & 1.1 & 8.3 $\pm$ 1.0 \\
2.0 $\pm$ 0.4 & +9.4 & 0.4 && $\leq$ 1.3 & \ldots & \ldots && $\leq$ 1.5 & 
\ldots & \ldots && $\leq$ 1.0 & \ldots & \ldots & $\leq$ 0.5 \\
0.9 $\pm$ 0.2 & +12.0 & \ldots && $\leq$ 1.3 & \ldots & \ldots && $\leq$ 1.5 & 
\ldots & \ldots && $\leq$ 1.0 & \ldots & \ldots & $\leq$ 1.1 \\
\hline
\multicolumn{16}{c}{HD 24076} \\
\hline
1.4 $\pm$ 0.7 & +5.0 & 1.8 && $\leq$ 2.1 & \ldots & \ldots && $\leq$ 1.3 & 
\ldots & \ldots && $\leq$ 1.7 & \ldots & \ldots & $\leq$ 12.1 \\
8.6 $\pm$ 0.8 & +7.2 & 1.6 && 4.7 $\pm$ 0.7 & +7.5 & 2.2 && $\leq$ 1.3 & 
\ldots & \ldots && 9.7 $\pm$ 0.6 & +6.9 & 1.1 & 11.3 $\pm$ 1.3 \\
\enddata
\tablenotetext{a}{From White et al. (2001).}
\tablenotetext{b}{Column densities of very closely spaced Na~{\tiny I} 
components were summed for comparison purposes. The velocity stated for summed 
components is a weighted average, however both $b$-values are listed.}
\end{deluxetable*}

\subsection{Total Equivalent Widths}

An extensive comparison of our values of total $W_{\lambda}$ with those found 
in the literature appears in Table 2 for the five species studied. A thorough 
examination reveals that our values, obtained by summing all of the individual 
components along a line of sight, are consistent with most previous 
determinations, many of which derive from only one resolved component. Of the 
37 spectra in our investigation with counterparts in the literature, 
significant discrepancies in total $W_{\lambda}$ (greater than 3-$\sigma$) are 
noted in only 6. Stellar contamination most likely affected the measurements 
by White (1984\emph{a}) of \ion{Ca}{2} toward 20 Tau and 22 Tau and of CH 
toward HD~23512. White's (1973) measurement of \ion{Ca}{2} toward $\eta$ Tau 
is also suspected to suffer from inadequate normalization of a stellar 
feature. Our value for $W_{\lambda}$(\ion{Ca}{2}) toward 27 Tau is 
substantially larger than the consistent values given by three previous 
investigations (Younan \& Dufton 1984; White 1984\emph{a}; Marschall \& Hobbs 
1972). This is due to our ability to detect weaker features in the wings of 
the more prominent absorption line. Similarly, the determination of 
$W_{\lambda}$(CH$^+$) toward HD~23753 by Frisch (1972) was affected by her 
detection limit of $\sim$ 6 m\AA. In the case of non-detections, since our 
upper limits are significantly lower than those of all preceding studies, our 
results further constrain the detectable amount of absorption from these less 
abundant species along the observed sight lines. We also point out that while 
Federman (1982) claimed a detection of CH toward 20 Tau, it is but a 
2.5-$\sigma$ detection and is consistent with our quoted upper limit. The 
average uncertainties in our measurements of $W_{\lambda}$ for individual 
components fall below 1 m\AA{} for all four detected species. Specifically, 
$\sigma_W$ = 0.7 m\AA{} for CN, 0.4 m\AA{} for \ion{Ca}{2}, and 0.3 m\AA{} for 
CH$^+$ and CH. Since the uncertainties from the literature are usually greater 
than 1 m\AA, our measurements are among the most precise now available.

\subsection{Column Densities}

Column densities for individual components were interpolated from curves of 
growth based on the measured equivalent widths and the parameters listed in 
Table 3. The adopted $b$-values, 1.6 km s$^{-1}$ for \ion{Ca}{2}, CH$^{+}$, 
and CH and 0.5 km s$^{-1}$ for CN, are average values for the respective 
species. However, most lines are weak so that measured equivalent widths for a 
given species fall on the linear portion of the curve of growth. Thus, the use 
of different $b$-values does not impact the derived column densities in any 
appreciable way. In particular, the largest optical depth at line center 
encountered under our assumptions is $\tau = 1.03$ for the R(0) line of CN 
toward HD 23512. Table 4 summarizes our results for the individual components 
of \ion{Ca}{2}~K, CH$^{+}$, and CH, listing the column density, $N$, radial 
velocity, $v_{\mathrm{\scriptstyle LSR}}$, and $b$-value of each component. 
The reported uncertainty in $N$ for a given component was inferred from the 
measured uncertainty in the value of $W_{\lambda}$. Table 4 also displays the 
relevant parameters of the components found by White et al. (2001) in their 
survey of interstellar \ion{Na}{1} in the Pleiades. The last column gives the 
\ion{Na}{1}/\ion{Ca}{2} column density ratio which will be discussed further 
in \S{} 4.2. For compactness, the CN column densities toward HD 23512 are not 
listed in Table 4 but are given here. $N$(CN) = 2.7 $\pm$ 0.3, 1.6 $\pm$ 0.3, 
and $\leq$ 2.8 in units of $10^{12}$ cm$^{-2}$ for the R(0), R(1), and P(1) 
lines, respectively. These quantities will be important later for deriving the 
density of the molecular cloud along this line of sight.

\tabletypesize{\scriptsize}
\begin{deluxetable}{lcccc}
\tablecolumns{5}
\tablewidth{0.4\textwidth}
\tablenum{5}
\tablecaption{Mean Velocities and Doppler Parameters}
\tablehead{ \colhead{Species} & \colhead{Stars} & \colhead{Components} & 
\colhead{$<$$v_{\mathrm{{\scriptscriptstyle LSR}}}$$>$} & \colhead{$<$$b$$>$} 
\\
\colhead{} & \colhead{} & \colhead{} & \colhead{(km s$^{-1}$)} & 
\colhead{(km s$^{-1}$)} }
\startdata
Ca~{\tiny II}\tablenotemark{a} & 20 & 65 & +6.2 & 1.5 $\pm$ 0.4 \\
 & 20 & 55 & +7.1 & 1.5 $\pm$ 0.4 \\
CH$^{+}$ & 19 & 23 & +8.6 & 1.8 $\pm$ 0.3 \\
CH & 6 & 6 & +8.7 & 1.6 $\pm$ 0.1 \\
CN & 1 & 1 & +9.4 & 0.5 $\pm$ 0.1 \\
\hline
Na~{\tiny I}\tablenotemark{b} & 20 & 42 & +7.9 & 1.2 $\pm$ 0.8 \\
 & 20 & 37 & +8.0 & 1.0 $\pm$ 0.3 \\
\enddata
\tablenotetext{a}{Second row is for positive velocity Ca~{\tiny II} 
components only.}
\tablenotetext{b}{From White et al. (2001). The second row excludes five broad 
Na~{\tiny I} components with $b>2.5$ km s$^{-1}$.}
\end{deluxetable}

\begin{figure*}
\centering
\includegraphics[width=0.7\textwidth]{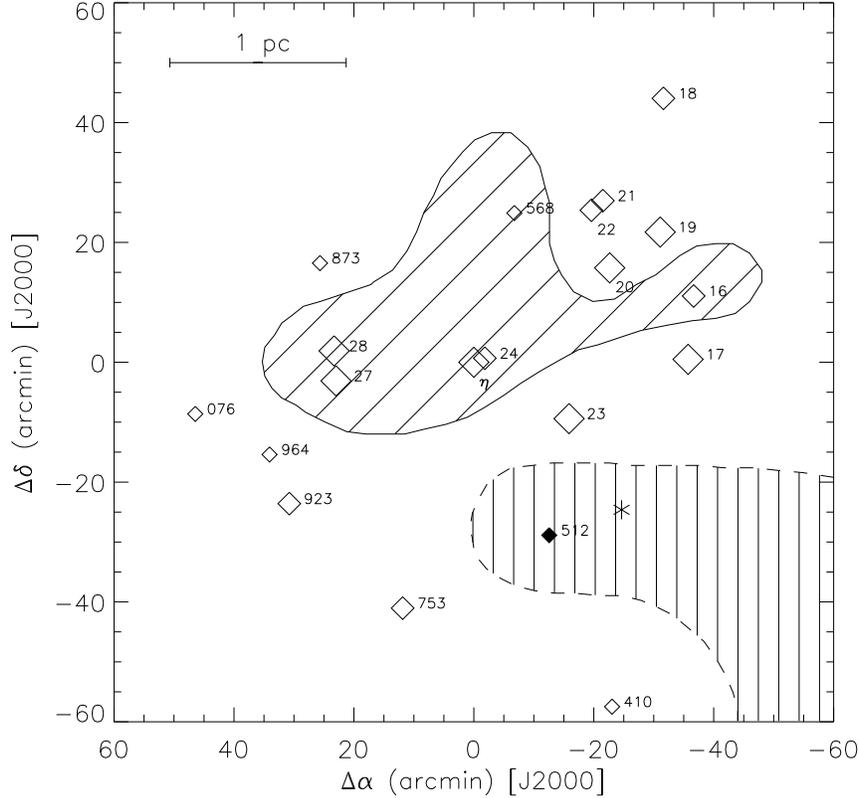}
\caption[Velocity components of CH.]{Contour plot of the observed velocity 
components of CH mapped onto the stars of the Pleiades. Solid symbols denote 
the strongest absorption components. The region enclosed by the solid line 
(diagonal hatching) has a velocity range (in km s$^{-1}$) of $+6.0 \leq 
v_{\mathrm{{\scriptscriptstyle LSR}}} \leq +7.5$ and the star enclosed by the 
dashed line (vertical hatching) has a velocity of +9.8 km s$^{-1}$. Although 
no component is detected toward 24 Tau, this star is included within the 
central contour because its column density upper limit of $\lesssim 0.8 
\times 10^{12}$ cm$^{-2}$ is consistent with a detection comparable in size 
to that found toward its binary companion, $\eta$ Tau, for which $N = 0.5 
\times 10^{12}$ cm$^{-2}$. The asterisk marks the position of peak CO emission 
from the molecular cloud discussed in the text.}
\end{figure*}

\subsection{Doppler Parameters and Radial Velocities}

The mean and dispersion in $b$-values were calculated for the detected species 
of this survey as well as for \ion{Na}{1} from the data provided by White et 
al. (2001). The results are displayed in Table 5. To facilitate a meaningful 
comparison, the values in the last row of the table were calculated by 
excluding five broad \ion{Na}{1} components with $b > 2.5$ km s$^{-1}$. The 
scatter in $b$-values for the species in our investigation is low and is 
comparable to the scatter found for the narrow \ion{Na}{1} components. Our 
mean $b$-values show good agreement with those from previous investigations, 
both within the Pleiades and in other regions of the ISM, for which spectra of 
comparable SNR were acquired. The ultra-high resolution survey of CH$^+$ and 
CH by Crane et al. (1995), for instance, included four Pleiades stars, 20 Tau, 
23 Tau, $\eta$ Tau, and 27 Tau. While they did not claim to detect CH along 
these sight lines, their mean CH$^+$ $b$-value for the four stars (1.9 $\pm$ 
0.5 km s$^{-1}$) agrees closely with the mean value for our sample of nineteen 
sight lines (1.8 $\pm$ 0.3 km s$^{-1}$). Welty et al. (1996) observed two 
Pleiades stars, 17 Tau and $\eta$ Tau, in a high-resolution survey of 
\ion{Ca}{2}. Their average $b$-value for the six components detected toward 
these two stars (1.5 $\pm$ 0.9 km s$^{-1}$) is identical to our average for 
twenty stars (1.5 $\pm$ 0.4 km s$^{-1}$), though the median value for their 
entire sample is somewhat smaller ($\sim$ 1.3 km s$^{-1}$). Pan et al. (2005) 
measured average $b$-values for a number of atomic and molecular species in an 
extensive examination of interstellar absorption in three star-forming 
regions, $\rho$ Oph, Cep OB2, and Cep OB3. Based on detections of 350 total 
components for \ion{Ca}{2} and over a hundred components for the other species 
excluding CN for which they detect 50 components, these authors found average 
$b$-values of 1.62 $\pm$ 0.28 km s$^{-1}$ for \ion{Ca}{2}, 1.96 $\pm$ 0.23 km 
s$^{-1}$ for CH$^+$, 1.04 $\pm$ 0.18 km s$^{-1}$ for CH, and 0.83 $\pm$ 0.11 
km s$^{-1}$ for CN. Given the uncertainties, the mean $b$-values in Table 5 
are consistent with these results, except in the case of CH. The larger value 
we obtain for this species (1.6 $\pm$ 0.1 km s$^{-1}$) may have implications 
for the environment responsible for the formation of these lines.

Previous studies have shown that CH can exist in both low-density gas 
($n\sim100$ cm$^{-3}$) where it is produced via non-thermal CH$^+$ chemistry 
(Draine \& Katz 1986; Zsarg\'o and Federman 2003), and in higher-density 
regions ($n\sim600$ cm$^{-3}$) where CN resides (Federman et al. 1994). CH 
components associated with CH$^+$ can be identified by a broad ``CH$^+$-like'' 
absorption profile, whereas those found in regions where CN is detected will 
exhibit a sharper ``CN-like'' profile (Lambert, Sheffer, \& Crane 1990). 
In the analysis of Pan et al. (2005), CN-like CH components were discovered to 
have an average $b$-value indistinguishable from that of CN components 
(0.90 $\pm$ 0.11 km s$^{-1}$), while CH$^+$-like CH components had a larger 
average $b$-value (1.10 $\pm$ 0.16 km s$^{-1}$). The CH components examined in 
the Pleiades are more closely associated with CH$^+$ than along average sight 
lines. Indeed, our mean CH $b$-value is nearly indistinguishable from our mean 
value for CH$^+$ and well above the $b$-values of the only detected CN lines 
(0.5 and 0.6 km s$^{-1}$ for R(0) and R(1), respectively, toward HD 23512). 
Toward most of the Pleiades, therefore, CH is found to be CH$^+$-like as 
opposed to CN-like. That is, CH is linked to the formation of CH$^+$ rather 
than CN and traces lower-density interstellar clouds. This conclusion would 
not be altered had we considered $\Lambda$-doubling in the CH line, which 
would lower our CH $b$-values by $\sim0.2$ km s$^{-1}$.

\begin{figure*}
\centering
\includegraphics[width=0.7\textwidth]{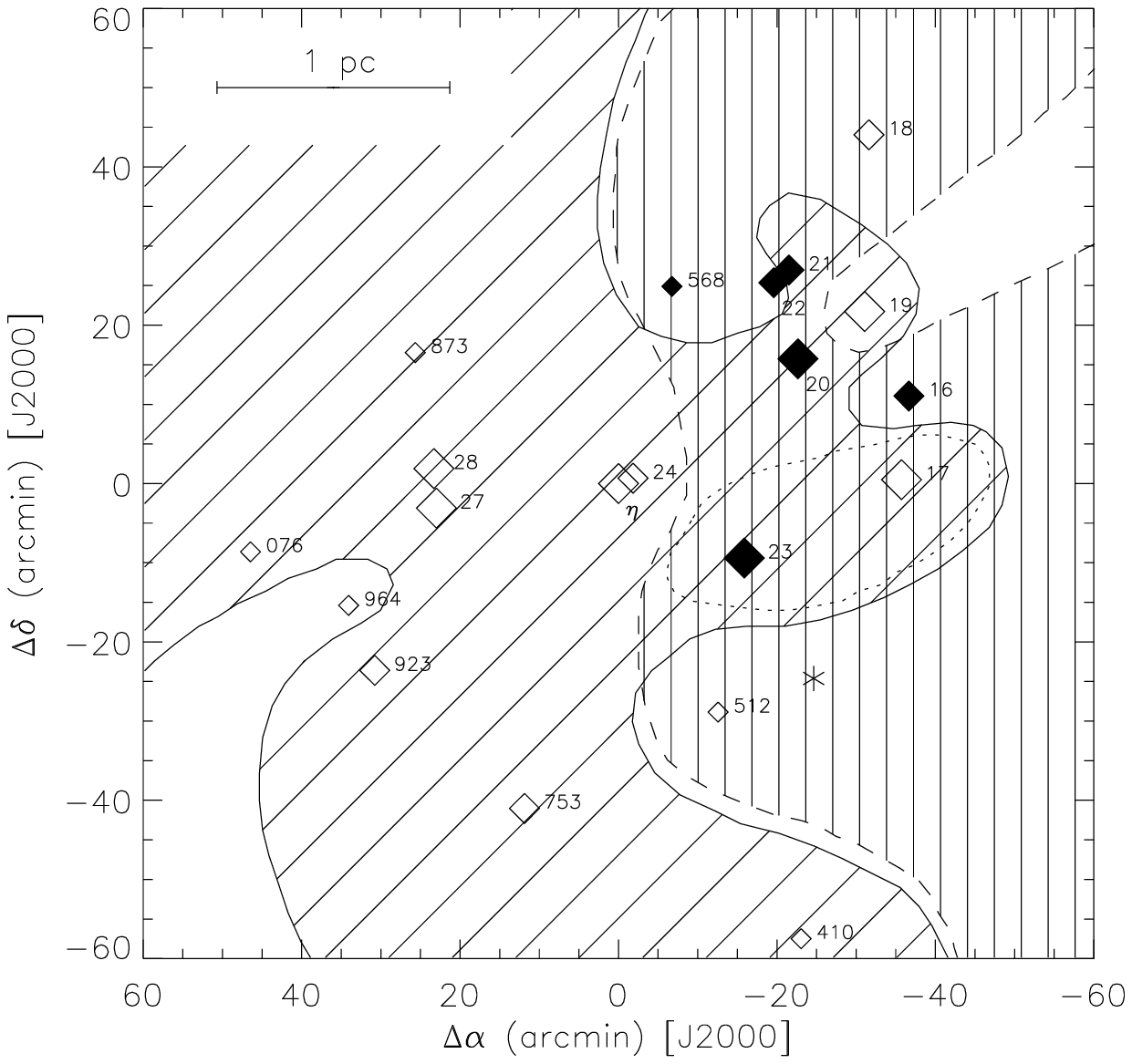}
\caption[Velocity components of CH$^{+}$.]{Contour plot of the observed 
velocity components of CH$^{+}$ mapped onto the stars of the Pleiades. Solid 
symbols denote the strongest absorption components. The area between the solid 
lines (diagonal hatching) has a velocity range (in km s$^{-1}$) of $+6.5 \leq 
v_{\mathrm{{\scriptscriptstyle LSR}}} \leq +8.1$ and between the dashed lines 
(vertical hatching) a velocity range of $+8.6 \leq 
v_{\mathrm{{\scriptscriptstyle LSR}}} \leq +10.3$. The two sight lines 
enclosed by the dotted line have velocities of +10.2 and +10.3 km s$^{-1}$. 
Again, the asterisk marks the position of peak CO emission from the molecular 
cloud.}
\end{figure*}

Table 5 also presents the mean radial velocities of the various species 
weighted by the column density of each component. These quantities yield 
information on the bulk kinematic properties of the various constituents of 
the interstellar gas near the Pleiades. The column-density weighted mean 
velocity for \ion{Ca}{2} in our sample is $\sim$ +6 km s$^{-1}$, whereas for 
molecular species this average is closer to $\sim$ +9 km s$^{-1}$. If only 
positive \ion{Ca}{2} velocity components are included, the mean velocity 
increases to $\sim$ +7 km s$^{-1}$. White (1984\emph{a}) was first to note the 
kinematic separation of atomic and molecular gas in this cluster. His results 
yielded average velocities of $\sim$ +7 km s$^{-1}$ for \ion{Ca}{2} and $\sim$ 
+8 km s$^{-1}$ for CH$^+$. The distinction may even be more pronounced in the 
average velocities reported here especially considering our measurements of CH 
and CN velocities. The \ion{Na}{1} components (White et al. 2001) have a 
weighted mean velocity of $\sim$ +8 km s$^{-1}$ in somewhat closer agreement 
with the molecular species than with \ion{Ca}{2}, suggesting that ionized and 
neutral gas may also be kinematically distinct. In the following section, the 
bulk motions described above are examined in greater detail by identifying 
individual velocity components present in the atomic and molecular gas.

\section{DATA ANALYSIS}

\subsection{Velocity Components}

In order to analyze the complex interactions between the interstellar gas and 
the stars of the Pleiades and to attempt to map the extent of the clouds 
involved in these interactions, contour maps were generated showing the 
spatial distribution of individual velocity components for species with 
detectable absorption along more than one line of sight. These maps are 
presented in Figures 6$-$8 for CH, CH$^+$, and \ion{Ca}{2} in order of 
increasing complexity. The contours in these figures were drawn by eye to 
include all sight lines which exhibit a given component. However, since we 
only have information at a limited number of discrete points, unlike radio 
data for instance, the boundaries shown are approximate at best. As a rule, if 
a component was found along one sight line but not along another adjacent to 
it, a boundary was drawn half way between. When no sight line was available to 
constrain the outer portions of a given component, contours were drawn to the 
edge of the plot.

\subsubsection{CH and CH$^+$ Velocity Components}

For the molecular species, CH and CH$^+$, one component was typically found 
per absorbing sight line, falling in either of two categories, those with 
$v_{\mathrm{\scriptstyle LSR}}$ $\sim$ +7 km s$^{-1}$ and those with 
$v_{\mathrm{\scriptstyle LSR}}$ $\sim$ +9.5 km s$^{-1}$. The former correspond 
to the ``central'' component in the \ion{Na}{1} profiles analyzed by White 
(2003) who identifies this absorption with the pervasive foreground gas 
thought to be associated with the Taurus dust clouds (see also White \& Bally 
1993). The latter, also detected in the \ion{Na}{1} data, are termed the 
``red'' component by White (2003) who suggests that this is either Taurus gas 
redshifted to higher radial velocities by the interaction with cluster stars 
or an independent gas cloud also interacting with the Pleiades. In our data, 
the strongest molecular absorption, including the only absorption detected in 
CN, is associated with the higher-velocity component. The lower-velocity gas 
is more pervasive and the two components generally occupy different regions of 
the cluster.

\begin{figure*}
\centering
\includegraphics[width=0.7\textwidth]{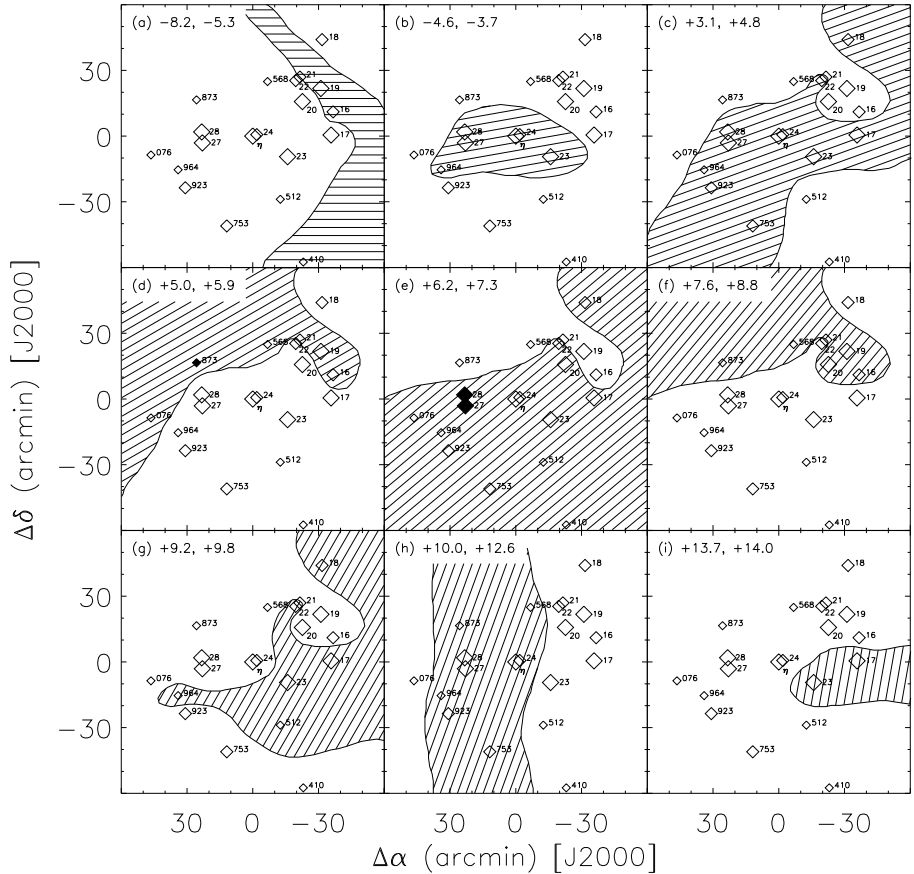}
\caption[Velocity components of Ca~{\scriptsize II}.]{Contour plots of the 
observed velocity components of Ca~{\scriptsize II} mapped onto the stars of 
the Pleiades. Solid symbols denote the strongest absorption components. The 
minimum and maximum velocities (in km s$^{-1}$) for each panel are labeled.}
\end{figure*}

Figure 6 clearly illustrates the spatial distinction between 7 km s$^{-1}$ and 
9.5 km s$^{-1}$ gas in the velocity components of CH and demonstrates the 
trend in absorption strength as well. There are five weak components with a 
mean velocity of +6.9 km s$^{-1}$ concentrated in the central region of the 
cluster. Four of these have column densities below $1\times10^{12}$ cm$^{-2}$ 
and the fifth, toward HD 23568, has a column density of $1.6\times10^{12}$ 
cm$^{-2}$. A much stronger component, with $N=5.4\times10^{12}$ cm$^{-2}$, is 
detected at a velocity of +9.8 km s$^{-1}$ toward HD 23512. The sight line to 
HD 23512 passes through a small molecular cloud first mapped in CO emission by 
R.S. Cohen (1975, unpublished; see Federman \& Willson 1984). The asterisk in 
Figure 6 marks the position of peak CO emission from the cloud. Federman and 
Willson published Cohen's CO data and added additional radio observations of 
CO and $^{13}$CO, as well as CH and OH, determining the velocity of the cloud 
to be about 10 km s$^{-1}$. Our velocity for the CH component toward HD 23512 
is consistent with their determination. In addition, the R(0) and R(1) lines 
of CN detected toward this star give velocities of +9.4 and +9.5 km s$^{-1}$, 
respectively. CN traces dense clouds and the extensive study by Pan et al. 
(2005) found that strong CH absorption is associated with CN (see \S{} 3.3). 
Thus, it is not surprising that CN is only found along a sight line containing 
a molecular cloud and that the CN-like CH component in this direction has a 
consistent velocity.

Although the CH$^+$ components in Figure 7 exhibit more complexity in their 
spatial distribution, the general trends are still apparent. Fourteen 
components are detected at a mean velocity of +7.3 km s$^{-1}$. These are 
distributed in mainly the central and eastern parts of the cluster, with some 
components extending to the south and to the west. The majority have column 
densities of $\sim3$$-$$5\times10^{12}$ cm$^{-2}$. The higher-velocity 
components responsible for the largest column densities of CH$^+$ are 
concentrated in the western half of the cluster. Nine components are found at 
a mean velocity of +9.3 km s$^{-1}$ and six of these have column densities 
greater than $15\times10^{12}$ cm$^{-2}$. The components toward 17 Tau and 23 
Tau have the highest velocities of this group, +10.3 and +10.2 km s$^{-1}$, 
respectively, noticeably higher than the velocity of any other component. 
These velocities coincide with the velocity of the molecular cloud discussed 
above and the sight lines toward these stars lie along the cloud's outer edge 
as seen in molecular emission (Federman \& Willson 1984). Both are strong 
indications that the material responsible for the absorption originated in the 
molecular cloud. Interestingly, no 7 km s$^{-1}$ gas is detected in the 
cloud's immediate vicinity (toward HD 23512) but is seen in weak absorption to 
the north (toward 17 Tau and 23 Tau) and in more moderate absorption to the 
south (toward HD 23410). We do not detect any systematic velocity shifts 
between our CH and CH$^+$ lines. The average difference in velocity for 
components detected in both species is $v$(CH$^+)-v$(CH$)=0.1$ km s$^{-1}$. 
Our CH velocities also agree well with those of \ion{Na}{1}, with an average 
difference for individual sight lines of $v$(CH$)-v$(\ion{Na}{1}$)=-0.1$ km 
s$^{-1}$.

\subsubsection{\ion{Ca}{2} Velocity Components}

While the molecular data can be adequately represented by two distinct 
velocity components, the situation for \ion{Ca}{2} is considerably more 
complex. Up to five separate components are identified along a single line of 
sight. In all, 55 components are detected with velocities distributed almost 
continuously from +3.1 to +14.0 km s$^{-1}$. Another 10 are found with 
velocities between $-$8.2 and $-$3.7 km s$^{-1}$. Velocities alone are not 
enough to categorize components with such diversity. Thus, the panels 
displayed in Figure 8 were created by classifying the \ion{Ca}{2} components 
not only according to their velocity, but also to their proximity to like 
sight lines and their overall position in the cluster. The notable features of 
this figure are as follows.

The most pervasive group of \ion{Ca}{2} components (Figure 8\emph{e}) has a 
mean velocity of +6.8 km s$^{-1}$, clearly associating it with the central 
component of White's (2003) analysis and that found in our molecular data. 
This component pervades the observed extent of the cluster with the exception 
of five sight lines along the northern edge, 16 Tau, 19 Tau, 21 Tau, HD 23568, 
and HD 23873. Unlike the weaker molecular feature at this velocity, the 
central \ion{Ca}{2} component is the dominant source of atomic absorption in 
the Pleiades with column densities ranging from $4$$-$$18\times10^{10}$ 
cm$^{-2}$. Proceeding outward in velocity from the central component, there 
are two groups (Figures 8\emph{d} and 8\emph{f}) which appear nearly identical 
in their morphology and spatial extent. These groups have mean velocities of 
+5.5 and +8.2 km s$^{-1}$, respectively, and seem to populate the apparent 
void in the central component. Indeed, all five of the aforementioned northern 
sight lines have components at both the slightly higher and slightly lower 
velocities. We interpret this bifurcation in velocity components, which 
appears smooth at our resolution, as evidence of a velocity gradient in the 
pervasive central cloud. However, the number of components belonging to groups 
8\emph{d} and 8\emph{f} may be complicated by unresolved component blending in 
the central component (8\emph{e}). For example, Welty et al. (1996) fit the 
asymmetric central region of their higher-resolution \ion{Ca}{2} profile 
toward $\eta$ Tau with two components with velocities of +6.2 and +7.7 km 
s$^{-1}$, compared to our one component with a velocity of +7.2 km s$^{-1}$. 
In our data, a hint of this asymmetry in the central region of the profile can 
be seen for $\eta$ Tau, as well as for 27 Tau and 28 Tau. In general, though, 
the $b$-values for these three groups of \ion{Ca}{2} components are quite 
similar ($\sim1.6$ km s$^{-1}$) and thus do not indicate that unresolved 
structure is a major concern.

Moving out still further in velocity from the central component, we encounter 
two additional groups (Figures 8\emph{c} and 8\emph{g}) with analogous spatial 
characteristics. These groups, with mean velocities of +3.8 and +9.4 km 
s$^{-1}$, respectively, exhibit identical structure in the cluster's northwest 
quadrant, but the lower-velocity components (8\emph{c}) are more extensive in 
the southeast. Interestingly, the six sight lines that are missing from the 
feature in Figure 8\emph{g} have components at even higher velocities (Figure 
8\emph{h}). This group (8\emph{h}) has a mean velocity of +11.5 km s$^{-1}$. 
Taken together, the components in Figures 8\emph{g} and 8\emph{h} are roughly 
comparable to the red component in White's (2003) classification scheme. 
Those in Figure 8\emph{c} correspond to what White (2003) terms the 
``blue'' component, with the implication that gas from the central cloud at 
$v_{\mathrm{\scriptstyle LSR}}$ $\sim$ +7 km s$^{-1}$ has been blueshifted by 
its interaction with the cluster to lower radial velocities. The spatial 
correlations and morphological similarities of redshifted and corresponding 
blueshifted components in Figure 8 are striking and support the claim that 
velocity gradients reflect the interaction on a cloud. Furthermore, the mean 
velocities found for the components of Figures 8\emph{c} and 8\emph{g} differ 
from the mean velocities of the previous two groups (Figures 8\emph{d} and 
8\emph{f}) by an amount consistent with the velocity differences between those 
groups and the central component. In other words, Figures 8\emph{c} through 
8\emph{g} present successive velocity intervals of approximately 1.4 km 
s$^{-1}$ evincing the symmetry of the interstellar gas in the Pleiades traced 
by \ion{Ca}{2} absorption.

The remaining panels of Figure 8 present further components which do not 
conform to the above symmetries. The components with the most extreme positive 
velocities (Figure 8\emph{i}) inhabit a region directly west of the cluster 
center. It is notable that the two sightlines showing this component, 17 Tau 
and 23 Tau with velocities of +14.0 and +13.7 km s$^{-1}$, respectively, are 
the same as those with the highest CH$^+$ velocities, although the actual 
velocities do not agree. The proximity of these highly redshifted components 
to the edge of the molecular cloud is suggestive of strong interactions 
between cloud material and the luminous blue giants. Figures 8\emph{a} and 
8\emph{b} reveal the distribution of \ion{Ca}{2} absorption detected at 
negative velocities. The six components concentrated in the central region of 
the cluster (8\emph{b}) have a mean velocity of $-$4.1 km s$^{-1}$ in close 
agreement with the negative-velocity components found in \ion{Na}{1}. White 
(2003) names this the ``shocked'' component, referring to the prediction that 
strongly blueshifted absorption features would arise in foreground gas shocked 
by the interaction with cluster stars. Four additional shocked components are 
detected along the cluster's western edge (8\emph{a}). These have even more 
strongly blueshifted velocities with a mean of $-$6.3 km s$^{-1}$. White finds 
shocked \ion{Na}{1} components toward only HD 23410, $\eta$ Tau, 27 Tau, and 
HD 23964 yet hints at unpublished \ion{Ca}{2} data confirming additional 
detections toward 21 Tau and 28 Tau. We also detect shocked \ion{Ca}{2} 
components toward 21 Tau and 28 Tau and add to this list 16 Tau, 19 Tau, 
23 Tau, and 24 Tau. The shocked atomic components provide an important 
observational constraint when considering the types of interactions needed to 
explain the features of the interstellar medium in the Pleiades.

\begin{figure}
\centering
\includegraphics[width=0.4\textwidth]{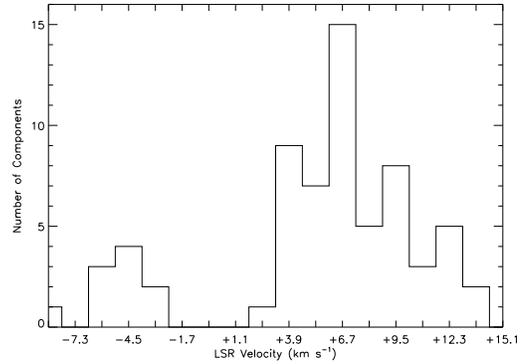}
\caption[Histogram of Ca~{\scriptsize II} velocity components.]{Representative 
histogram of Ca~{\scriptsize II} velocity components with a bin size of 
1.4 km s$^{-1}$, comparable to our velocity resolution.}
\end{figure}

We performed a statistical analysis to determine whether binning the 
\ion{Ca}{2} velocity components in Figure 8 differently would severely alter 
our results. To this end, a series of histograms were constructed with 
successive bin sizes from 0.2 km s$^{-1}$ to 1.9 km s$^{-1}$. In almost every 
case, four distinct peaks could be seen for the positive-velocity components, 
with another single peak evident for the components with negative velocity. A 
representative histogram is shown in Figure 9 with a bin size of 1.4 km 
s$^{-1}$, comparable to the resolution in our data. The peaks in the figure 
are centered about values of $-$4.5, +3.9, +6.7, +9.5, and +12.3 km s$^{-1}$, 
closely echoing the mean velocities found for the components in Figures 
8\emph{b}, 8\emph{c}, 8\emph{e}, 8\emph{g}, and 8\emph{h}, respectively. 
Because variations in bin size have a negligible effect on the results 
presented in Figure 8, these features are not an artifact of a particular 
classification scheme, but represent physical structure in the interstellar 
atomic gas.

\subsection{\ion{Na}{1}/\ion{Ca}{2} Column Density Ratios}

The last column of Table 4 displays the $N$(\ion{Na}{1})/$N$(\ion{Ca}{2}) 
ratios for our \ion{Ca}{2} components common to the \ion{Na}{1} investigation 
of White et al. (2001) and upper limits for \ion{Ca}{2} components not 
detected in \ion{Na}{1}. Figure 10 plots these ratios as a function of 
$v$(\ion{Ca}{2}). Routly and Spitzer (1952) were first to demonstrate that the 
\ion{Na}{1}/\ion{Ca}{2} column density ratios are inversely proportional to 
the velocity of the material. This phenomenon, known as the Routly-Spitzer 
effect, was later confirmed by Siluk and Silk (1974), who found that the 
decrease in $N$(\ion{Na}{1})/$N$(\ion{Ca}{2}) was most evident for components 
with absolute velocities greater than $\sim$ 20 km s$^{-1}$. The 
Routly-Spitzer effect is generally attributed to variations in the gas-phase 
abundance of calcium in clouds at different velocities. Calcium is depleted 
onto the surfaces of dust grains in low-velocity diffuse interstellar clouds 
resulting in high \ion{Na}{1}/\ion{Ca}{2} ratios in these environments. 
Heating from shock waves and cloud collisions in higher-velocity material can 
cause grain destruction which releases calcium back to its gaseous phase, 
yielding a lower \ion{Na}{1}/\ion{Ca}{2} ratio. The data in Figure 10 show an 
indication of this trend but there are not enough high-velocity detections to 
be certain.

\begin{figure}
\centering
\includegraphics[width=0.4\textwidth]{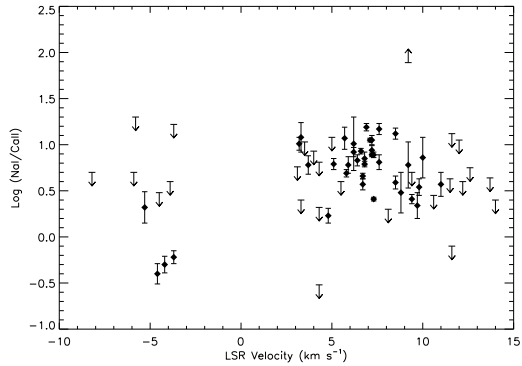}
\caption[Na~{\scriptsize I}/Ca~{\scriptsize II} column density ratios.]
{$N$(Na~{\scriptsize I})/$N$(Ca~{\scriptsize II}) ratios as a function of 
$v$(Ca~{\scriptsize II}). Of the four data points with 
$v_{\mathrm{{\scriptscriptstyle LSR}}}<0$, three have ratios less than one. 
These are for the shocked components toward $\eta$ Tau, 27 Tau, and HD 23964. 
The fourth, with a ratio of $\sim$ 2, is for the shocked component toward HD 
23410. The lower limit near the top of the figure is for the +9.2 km s$^{-1}$ 
component toward HD 23512.}
\end{figure}

More importantly, because calcium depletion is assumed to be 
density-dependent, our \ion{Na}{1}/\ion{Ca}{2} ratios can be used to 
characterize the physical conditions of the interstellar clouds responsible 
for the detected absorption, providing a test for our derived column 
densities. The majority of positive-velocity components in Figure 10 have 
$N$(\ion{Na}{1})/$N$(\ion{Ca}{2}) of order a few to 10, consistent with the 
values typical of diffuse, non-molecular clouds with densities of 
$\sim10$ cm$^{-3}$ (Hobbs 1976). The only exception to this is the 
+9.2 km s$^{-1}$ component toward HD 23512. The \ion{Na}{1}/\ion{Ca}{2} ratio 
for this component (the lower limit near the top of Figure 10) is of order 
100, more characteristic of a denser, molecular cloud where calcium is 
depleted onto grain surfaces more effectively (Crawford et al. 1989; 
Crawford 1992). In contrast, the +6.2 km s$^{-1}$ component along this sight 
line has a ratio of $\sim$ 10 placing it among the diffuse components where 
calcium depletion is less severe. Since these results predict the proper 
relationship between velocity components and the expected physical 
characteristics of the associated clouds, it gives us confidence in the 
accuracy of our derived column densities. A conspicuous feature of Figure 10 
is that the \ion{Na}{1}/\ion{Ca}{2} ratios for shocked atomic components 
(those with $v_{\mathrm{\scriptstyle LSR}} < 0$) are less than one except 
toward HD 23410 for which $N$(\ion{Na}{1})/$N$(\ion{Ca}{2}) $\sim$ 2. In the 
environment near the Pleiades, a ratio less than one probably implies that 
photoionization from bright cluster stars has led to a reduction in neutral 
sodium. This effect would be less pronounced for gas near HD 23410 lying far 
from the more luminous stars and producing little luminosity of its own. The 
shocks themselves do not have high enough velocities to contribute much 
gas-phase calcium from grain processing.

\subsection{Physical Conditions}

It is possible to extract the physical conditions of the interstellar medium 
near the Pleiades directly from our absorption-line data. In this section, we 
derive estimates for $n$, the total gas density of the absorbing clouds, and 
$I_{uv}$, the enhancement of the UV radiation field due to cluster stars over 
the average Galactic field, by adopting various models of diffuse-cloud 
chemistry.

\subsubsection{Steady-State Model of CH Formation from CH$^+$}

For the majority of sight lines in our sample, CH is found to be closely 
associated with CH$^+$. In such environments, the total gas density, defined 
as $n=n$(H) + $2n$(H$_{2}$), may be obtained by studying the reactions which 
lead to the production of CH from CH$^+$. In the calculations that follow, we 
do not explicitly consider the factors responsible for CH$^+$ formation. 
Instead, we follow the chemistry once CH$^+$ is present in appreciable 
amounts. In particular, we use the observed column density of CH$^+$ to 
determine the gas density implied by the amount of CH present. Assuming that 
the most important reactions

\begin{displaymath}
\begin{array}{c}
\mathrm{CH}^++\mathrm{H}_2\to\mathrm{CH}_2^++\mathrm{H},\\
\mathrm{CH}_2^++\mathrm{H}_2\to\mathrm{CH}_3^++\mathrm{H},\textrm{ and}\\
\mathrm{CH}_3^++\mathrm{e}\to\mathrm{CH}+\mathrm{H}_2\textrm{ or }2\mathrm{H}
\end{array}
\end{displaymath}

\noindent
satisfy the conditions for a steady state, the rate equation for CH may be 
written (Welty et al. 2006)

\begin{equation}
N(\mathrm{CH})=\frac{0.67\;k(\mathrm{CH}^+,\mathrm{H}_2)\;N(\mathrm{CH}^+)\;
f(\mathrm{H}_2)\;n}{2\;I_{uv}\;G(\mathrm{CH})},
\end{equation}

\noindent
where the factor 0.67 represents the fraction of dissociative recombinations 
of CH$_3^+$ producing CH (Herbst 1978; Vejby-Christensen et al. 1997), 
$k$(CH$^+$,H$_2) = 1.2 \times 10^{-9}$ cm$^3$ s$^{-1}$ is the rate coefficient 
for the initial reaction (Draine \& Katz 1986), $f$(H$_2) = 2 
N$(H$_2$)/$N_{tot}$(H) is the fraction of hydrogen in molecular form, and 
$G$(CH$) = 1.3 \times 10^{-9} \mathrm{e}^{-\tau_{uv}}$ s$^{-1}$ is the CH 
photodissociation rate for the average Galactic field (Federman \& Huntress 
1989). In the expression for $f$(H$_2$), $N_{tot}$(H) = $N$(\ion{H}{1}) + $2 
N$(H$_2$) denotes the total hydrogen column density along the line of sight. 
Eqn. (1) may be inverted to derive an expression for the gas density as a 
function of the CH/CH$^+$ column density ratio,

\begin{equation}
n=\frac{N(\mathrm{CH})}{N(\mathrm{CH}^+)}\;\frac{2\;I_{uv}\;
G(\mathrm{CH})}{0.67\;k(\mathrm{CH}^+,\mathrm{H}_2)\;f(\mathrm{H}_2)}.
\end{equation}

\tabletypesize{\scriptsize}
\begin{deluxetable*}{lccccccccccc}
\tablecolumns{11}
\tablewidth{0.8\textwidth}
\tablenum{6}
\tablecaption{Physical Conditions}
\tablehead{ \colhead{} & \multicolumn{4}{c}{From CH/CH$^{+}$ Column Density 
Ratio} & \colhead{} & \multicolumn{5}{c}{From H$_{2}$ Column Densities} \\
\cline{2-5} \cline{7-11} \\
\colhead{Star} & \colhead{$\tau_{uv}$} & 
\colhead{$N(\mathrm{CH})/N(\mathrm{CH}^{+})$} & 
\colhead{$I_{uv}$\tablenotemark{a}} & \colhead{$n$\tablenotemark{b}} & 
\colhead{} & \colhead{$N(J=0)$\tablenotemark{c}} & 
\colhead{$N(J=1)$\tablenotemark{c}} & \colhead{$N(J=4)$\tablenotemark{d}} & 
\colhead{$n$\tablenotemark{e}} & \colhead{$I_{uv}$\tablenotemark{f}} \\ 
\colhead{} & \colhead{} & \colhead{} & \colhead{} & \colhead{(cm$^{-3}$)} & 
\colhead{} & \colhead{($10^{19}$ cm$^{-2}$)} & \colhead{($10^{19}$ cm$^{-2}$)} 
& \colhead{($10^{15}$ cm$^{-2}$)} & \colhead{(cm$^{-3}$)} & \colhead{} \\
\colhead{(1)} & \colhead{(2)} & \colhead{(3)} & \colhead{(4)} & \colhead{(5)} 
& \colhead{} & \colhead{(6)} & \colhead{(7)} & \colhead{(8)} & \colhead{(9)} & 
\colhead{(10)} }
\startdata
16 Tau & 0.62 & $\geq$ 0.3\phn & 28.9 & $\geq$ 168 && \ldots & \ldots & 
\ldots & \ldots & \ldots \\
 & & $\leq$ 0.06 & & \phn$\leq$ 28 && \ldots & \ldots & \ldots & \ldots & 
\ldots \\
17 Tau & 0.31 & $\leq$ 0.8\phn & 21.9 & $\leq$ 404 && \ldots & \ldots & 
\ldots & \ldots & \ldots \\
 & & $\leq$ 0.5\phn & & $\leq$ 280 && \ldots & \ldots & \ldots & \ldots & 
\ldots \\
18 Tau & 0.31 & $\leq$ 0.3\phn & 13.3 & \phn$\leq$ 88 && \ldots & \ldots & 
\ldots & \ldots & \ldots \\
19 Tau & 0.25 & $\leq$ 0.2\phn & 34.5 & $\leq$ 158 && \ldots & \ldots & 
\ldots & \ldots & \ldots \\
20 Tau & 0.43 & $\leq$ 0.05 & 14.9 & \phn$\leq$ 16 && 2.4 $\pm$ 1.0 & 3.2 
$\pm$ 1.3 & 1.3 $\pm$ 0.2 & 89 $\pm$ 34 & 8.6 $\pm$ 3.4 \\
 & & $\leq$ 0.05 & & \phn$\leq$ 16 && \ldots & \ldots & \ldots & \ldots & 
\ldots \\
21 Tau & 0.43 & $\leq$ 0.3\phn & 14.7 & \phn$\leq$ 82 && \ldots & \ldots & 
\ldots & \ldots & \ldots \\
 & & $\leq$ 0.07 & & \phn$\leq$ 21 && \ldots & \ldots & \ldots & \ldots & 
\ldots \\
22 Tau & 0.37 & $\leq$ 0.06 & \phn7.9 & \phn$\leq$ 10 && \ldots & \ldots & 
\ldots & \ldots & \ldots \\
23 Tau & 0.62 & $\leq$ 0.5\phn & 20.5 & $\leq$ 192 && 7.8 $\pm$ 2.0 & 5.2 
$\pm$ 1.3 & 2.2 $\pm$ 0.2 & 48 $\pm$ 12 & 7.1 $\pm$ 1.8 \\
 & & $\leq$ 0.04 & & \phn$\leq$ 14 && \ldots & \ldots & \ldots & \ldots & 
\ldots \\
HD 23568 & 0.43 & $\geq$ 0.5\phn & \phn6.4 & \phn$\geq$ 72 && \ldots & 
\ldots & \ldots & \ldots & \ldots \\
 & & $\leq$ 0.06 & & \phn\phn$\leq$ 8 && \ldots & \ldots & \ldots & \ldots & 
\ldots \\
$\eta$ Tau & 0.25 & \phm{$\leq$} 0.2\phn & \phn8.8 & 46 $\pm$ 9 && 1.6 
$\pm$ 0.4 & 1.9 $\pm$ 0.5 & 0.8 $\pm$ 0.1 & 83 $\pm$ 20 & 6.3 $\pm$ 1.6 \\
HD 23753 & 0.25 & $\leq$ 0.2\phn & 11.0 & \phn$\leq$ 53 && \ldots & \ldots & 
\ldots & \ldots & \ldots \\
27 Tau & 0.25 & \phm{$\leq$} 0.1\phn & \phn6.1 & 16 $\pm$ 6 && \ldots & 
\ldots & \ldots & \ldots & \ldots \\
28 Tau & 0.19 & \phm{$\leq$} 0.1\phn & 10.3 & 40 $\pm$ 6 && \ldots & 
\ldots & \ldots & \ldots & \ldots \\
HD 23923 & 0.37 & $\leq$ 0.4\phn & \phn3.6 & \phn$\leq$ 29 && \ldots & 
\ldots & \ldots & \ldots & \ldots \\
\enddata
\tablenotetext{a}{UV flux calculated from parameters in White (1984\emph{b}).}
\tablenotetext{b}{Gas density calculated from parameters in columns (2) 
through (4). Two entries indicate multiple components along the line of sight.}
\tablenotetext{c}{H$_{2}$ column densities from Savage et al. (1977).}
\tablenotetext{d}{H$_{2}$ column density from Spitzer, Cochran, \& Hirshfeld 
(1974) for 20 Tau and Frisch \& Jura (1980) for 23 and $\eta$ Tau.}
\tablenotetext{e}{Gas density calculated from parameters in columns (6) 
through (8).}
\tablenotetext{f}{UV flux calculated from parameters in columns (6) through 
(8).}
\end{deluxetable*}

To estimate the gas density from eqn. (2), a few important parameters must 
first be determined, namely $f$(H$_2$) and $I_{uv}$. The first of these, the 
molecular fraction, is not well constrained in the vicinity of the Pleiades 
due to a dearth of reliable measurements of $N$(\ion{H}{1}). As a starting 
point in our analysis, we adopted a value of $f$(H$_2) = 0.1$ due to the 
strong presence of CH$^+$ in these clouds. A greater fraction of hydrogen in 
molecular form would rapidly destroy CH$^+$ by the reaction CH$^+$ + H$_2$ 
$\to$ CH$_2^+$ + H. Since $n$ is inversely proportional to $f$, a lower value 
of $f$ would increase the densities obtained from eqn. (2), a fact which is 
exploited in \S{} 4.3.2. Individual values of $I_{uv}$ are calculated using 
the following expression modified from eqn. (2) of White (1984\emph{b}),

\begin{equation}
I_{uv}=\left[1+\frac{\chi_{\lambda}(1)}{x^2}\right]\mathrm{e}^{-\tau_{uv}}+
\chi_{\lambda,c},
\end{equation}

\noindent
where $\chi_{\lambda}(1)$ is the ratio of the flux due to a specific star at a 
distance of 1 pc to that produced by the unattenuated galactic background, 
$\chi_{\lambda,c}$ is the corresponding ratio for the flux due to the other 
cluster stars at that position, and $x$ is the distance from the star in pc. 
The radiative parameters for 14 of our 20 stars (at $\lambda=1000$ \AA) are 
obtained from Table 1 of White (1984\emph{b}) from which $x$ is taken to be 
$x_{\mathrm{min}}$, which is the minimum distance at which radiation pressure 
is balanced by the drag force on dust grains slipping through the gas as it 
approaches the cluster. Yet, since $x$ in eqn. (3) refers to the gas 
responsible for the absorption, the adoption of these values assumes a close 
coupling between the gas and dust, though the actual distances used 
($x\sim0.1$$-$$0.4$ pc) are quite similar to those found in other analyses. 
For example, Jura (1977) argued for a distance of 0.1 pc for the gas toward 20 
Tau and Federman (1982) determined the distance to be 0.3 pc. The value 
obtained from White (1984\emph{b}) for the gas toward this star is also 0.3 
pc. Finally, to calculate $I_{uv}$ as well as $G$(CH), we approximate 
$\tau_{uv}$, the grain optical depth at 1000 \AA, by $2 \times A_V$ (Federman 
et al. 1994), determined from the reddening values in Table 1.

Table 6 displays the relevant parameters for each sight line and the resulting 
density estimates. Because the majority of sight lines show no detectable CH 
absorption, most of the densities in column (5) are only upper limits. We do 
not attempt to apply this method to the sight line toward HD 23512 because the 
assumptions leading to eqn. (2) would not be appropriate for the gas in a 
molecular cloud where CH is linked to CN chemistry. Of the remaining five 
sight lines with both CH$^+$ and CH in detectable amounts, two (16 Tau and HD 
23568) do not have components at similar velocities. Thus, only three actual 
estimates for density can be obtained from our data using this procedure. The 
values, $n$ = 46, 16, and 40 cm$^{-3}$ for the sight lines toward $\eta$ Tau, 
27 Tau, and 28 Tau, respectively, are significantly below the densities 
predicted by earlier investigations. The atomic ionization models of White 
(1984\emph{a}) favored a density of 400 cm$^{-3}$, while his chemical models 
of CH and CH$^+$ commonly employed values of 100 and 300 cm$^{-3}$ 
(White 1984\emph{b}). The 21 cm data examined by Gordon \& Arny (1984) also 
suggested a hydrogen density of $\sim 100$ cm$^{-3}$. However, a recent 
analysis by Zsarg\'o and Federman (2003) of \ion{C}{1} excitation yielded a 
gas density upper limit toward $\eta$ Tau, $n \leq 3.5$ cm$^{-3}$, an order of 
magnitude below our determination.

\subsubsection{Ultraviolet-Pumping Model of H$_2$ Rotational Excitation}

An alternative approach to obtaining the physical conditions of the diffuse 
clouds near the Pleiades involves the distribution of H$_2$ column densities 
in the $J$ = 0, 1, and 4 rotational levels of the ground vibrational state. 
Much of this analysis, adapted from Lee et al. (2002), is derived from the 
work of Jura (1974, 1975) who showed that the $J$ = 4 and 5 rotational levels 
of H$_2$ are populated primarily by photon pumping. In environments of low to 
moderate density, these levels will be depopulated by spontaneous emission. 
Thus, the relative populations of higher and lower rotational levels should 
indicate the density of the gas. While there may be other sources of 
excitation active in these clouds, such as collisions in shocked gas, the 
present model considers an extreme case in which all rotational excitation is 
due to UV pumping. Here, again, we assume steady-state conditions and obtain 
an expression for the gas density from eqn. (3) of Lee et al. (2002) adopting 
an H$_2$ formation rate coefficient, $R = 3 \times 10^{-17}$ cm$^3$ s$^{-1}$, 
appropriate for Galactic clouds. The expression is

\begin{equation}
n=9.2\times10^7\frac{N(4)}{N_{tot}(\mathrm{H})}\left\{\frac{0.26\;N(0)}{0.11\;
[N(0)+N(1)]}+0.19\right\}^{-1},
\end{equation}

\noindent
where $N(J)$ is the H$_2$ column density in rotational level $J$ and the total 
hydrogen column density is found from $N_{tot}$(H$) = 2 N$(H$_2)/f$(H$_2) = 
2\;[N(0) + N(1)]/f$(H$_2)$. In this case, therefore, $n$ is directly 
proportional to $f$. Column (9) of Table 6 presents the density estimates 
derived from eqn. (4) for the three sight lines with H$_2$ column densities 
measured by \emph{Copernicus} (Spitzer, Cochran, \& Hirshfeld 1974; Savage et 
al. 1977; Frisch \& Jura 1980). Unfortunately, only one of the sight lines, 
that toward $\eta$ Tau, has a corresponding value from the $N$(CH)/$N$(CH$^+$) 
method. Nevertheless, the densities obtained from eqn. (4), $n$ = 89, 48, and 
83 cm$^{-3}$ for 20 Tau, 23 Tau, and $\eta$ Tau, respectively, are 
consistently larger than those from the previous method by about a factor of 
two. Here we exploit the opposite dependence of $n$ on $f$ exhibited in eqns. 
(2) and (4) to constrain both the density and the molecular fraction. Table 7 
shows the effect on $n$ calculated from the two methods as $f$ is decreased 
slightly from 0.1 to 0.05. The analysis finds that a value of $f=0.07$ 
provides the best agreement among the various density estimates, a value 
consistent with the average value of $f$ for sight lines in Scorpius (Savage 
et al. 1977). The density toward $\eta$ Tau for this value of $f$ is then 
$n=62$ cm$^{-3}$.

\tabletypesize{\scriptsize}
\begin{deluxetable}{lccccccc}
\tablecolumns{8}
\tablewidth{0.5\textwidth}
\tablenum{7}
\tablecaption{Density as a Function of $f$(H$_2$)}
\tablehead{ \colhead{} & \multicolumn{3}{c}{$n$ (cm$^{-3}$) from CH/CH$^{+}$} 
& \colhead{} & \multicolumn{3}{c}{$n$ (cm$^{-3}$) from H$_{2}$} \\
\cline{2-4} \cline{6-8} \\
\colhead{Star} & \colhead{$f=0.10$} & \colhead{$f=0.07$} & \colhead{$f=0.05$} 
& \colhead{} & 
\colhead{$f=0.10$} & \colhead{$f=0.07$} & \colhead{$f=0.05$} \\  }
\startdata
20 Tau & $\leq$ \phn16 & $\leq$ \phn23 & $\leq$ \phn32 && 89 $\pm$ 34 & 62 
$\pm$ 24 & 44 $\pm$ 17 \\
 & $\leq$ \phn16 & $\leq$ \phn22 & $\leq$ \phn31 && \ldots & \ldots & \ldots 
\\
23 Tau & $\leq$ 192 & $\leq$ 274 & $\leq$ 384 && 48 $\pm$ 12 & 34 $\pm$ \phn8 
& 24 $\pm$ \phn6 \\
 & $\leq$ \phn14 & $\leq$ \phn19 & $\leq$ \phn27 && \ldots & \ldots & \ldots 
\\
$\eta$ Tau & 46 $\pm$ 9 & 66 $\pm$ 14 & 92 $\pm$ 19 && 83 $\pm$ 20 & 58 $\pm$ 
14 & 41 $\pm$ 10 \\
27 Tau & 16 $\pm$ 6 & 22 $\pm$ \phn9 & 31 $\pm$ 13 && \ldots & \ldots & \ldots 
\\
28 Tau & 40 $\pm$ 6 & 56 $\pm$ \phn8 & 79 $\pm$ 12 && \ldots & \ldots & \ldots 
\\
\enddata
\end{deluxetable}

Lee et al. (2002) derive another expression which relates the gas density to 
the ambient UV radiation field. We can use this relation to determine $I_{uv}$ 
based on the H$_2$ column densities in Table 6. From eqn. (4) of Lee et al. 
(2002) and using the above value for $R$,

\begin{equation}
I_{uv}=\frac{7.1\times10^{-23}\;n}{0.11\;\beta_0}\left[\frac{N_{tot}
(\mathrm{H})}{2\;f(\mathrm{H}_2)}\right]^{1/2},
\end{equation}

\noindent
where $\beta_0 = 5 \times 10^{-10}$ s$^{-1}$ is the H$_2$ photoabsorption rate 
corresponding to the average Galactic field and the factor 0.11 represents the 
fraction of absorptions leading to H$_2$ dissociation. The $I_{uv}$ values 
calculated from eqn. (5) are given in the last column of Table 6. While there 
are only three determinations, each is smaller by approximately a factor of 
two compared to the values in column (4) calculated from the parameters in 
White (1984\emph{b}). However, since eqn. (5) has no explicit $x$-dependence, 
we can determine what value of $x$ brings our $I_{uv}$ values into agreement. 
Solving eqn. (3) for $x$ and using $I_{uv}$ from column (10) of Table 6 yields 
$x$ = 0.9, 1.1, and 1.3 pc for 20 Tau, 23 Tau, and $\eta$ Tau, respectively. 
These are to be compared to the distances obtained from White (1984\emph{b}) 
for the same three stars, $x$ = 0.30, 0.32, and 0.72 pc, the values initially 
used in eqn. (3) to calculate the $I_{uv}$ values in column (4) of Table 6. By 
substituting eqn. (4) for $n$ into eqn. (5), it can be shown that this formula 
for $I_{uv}$ is also independent of $f$(H$_2$). Thus, if the larger distances 
just calculated were used in the comparative density analysis above, our 
best-match value of $f$ would decrease slightly to 0.06, thereby decreasing 
the density toward $\eta$ Tau to 52 cm$^{-3}$.

Despite the generally good agreement found from the above chemical models, the 
solutions are not unique and the number of free parameters certainly does not 
eliminate the need for further refinement. One example would be the 
application of a time-dependent model of CH$^+$ formation in this environment. 
Assuming a flow rate for the gas with respect to the cluster of 10 km s$^{-1}$ 
(Gordon \& Arny 1984) and a cloud thickness of 0.1 pc (White 1984\emph{a}), 
the timescale for cloud-cluster interactions is $\sim10^4$ yrs. A 
corresponding timescale for H$_2$ dissociation can be approximated by 
1/$G$(H$_2$), where $G$(H$_2) \simeq 0.11\beta_0$ is the H$_2$ 
photodissociation rate given above. If self-shielding is also included, the 
dissociation rate would be lower by at least a factor of 10 yielding a 
characteristic timescale of $\sim6\times10^3$ yrs. Because the respective 
timescales are comparable, we cannot decouple the two processes and 
time-dependent effects must ultimately be incorporated into any chemical model 
adopted to derive physical conditions.

\subsubsection{Diffuse-Cloud Model for CH and CN}

Following the analysis of CH and CN chemistry by Pan et al. (2005), we 
obtained a value for the gas density of the Pleiades molecular cloud toward HD 
23512. The model is appropriate for a diffuse cloud with large molecular 
content but does not include dark-cloud chemistry. As argued by Federman \& 
Willson (1984), the CO column density from the densest, most opaque portion of 
the cloud, $N$(CO$)\sim4.5\times10^{17}$ cm$^{-2}$, while larger than the 
values measured in diffuse gas, is significantly lower than those measured 
toward the centers of dark clouds. The Pleiades cloud is not, therefore, an 
intermediate zone between diffuse and dark clouds but simply a much less 
opaque molecular cloud than is normally observed in the Taurus region. Using 
our measured CH and CN column densities toward HD 23512 and setting $I_{uv}=1$ 
due to the position of this star almost a parsec distant (in projected 
separation) from the luminous cluster stars, the model suggests $n > 1600$ 
cm$^{-3}$. The original study by Federman \& Willson (1984) predicted $n \sim$ 
300$-$500 cm$^{-3}$ based on their $^{13}$CO data and the large velocity 
gradient model of Goldsmith et al. (1983). Bally and White (1986, unpublished; 
see White 2003) found a density of 300 cm$^{-3}$ within 10$^{\prime}$ of the 
position of peak CO emission and 800 cm$^{-3}$ at the position of the peak. 
The consistency of these densities with those derived by Federman \& Willson 
(1984) leads us to conclude that our lower limit is in error, being a factor 
of two larger than even the above peak value. Evidently, there is more CN 
along this line of sight than can be attributed solely to diffuse-cloud 
chemistry. The chemical reaction network common to dark clouds appears to 
contribute to CN production despite the unusually low density of this diffuse 
molecular cloud.

\section{DISCUSSION}

In \S{} 4.1, the relatively straightforward distribution of molecular gas in 
the Pleiades as traced by the molecules of CH$^+$, CH, and CN is contrasted 
with the rather complicated structures seen in the ionized atomic gas traced 
by \ion{Ca}{2}. The largest columns of molecular gas are associated with a 
velocity component at $v_{\mathrm{\scriptstyle LSR}}$ $\sim$ +9.5 km s$^{-1}$ 
concentrated in the western region of the cluster. Clearly, some of this 
material is related to the molecular cloud seen along the line of sight toward 
HD 23512, particularly that detected in CH and CN absorption in this 
direction. Indeed, the velocities of the CH and CN components here are 
consistent with the cloud velocity determined from molecular line emission by 
Federman \& Willson (1984) and, although the density of the cloud derived from 
our CH and CN column densities ($n > 1600$ cm$^{-3}$) is significantly larger 
than their predicted value ($\sim$ 300$-$500 cm$^{-3}$), the inclusion of 
dark-cloud chemistry should correct this discrepancy. Still, whether or not 
all molecular components that share the cloud velocity have a common origin 
remains elusive. The general conclusions of the expansive study by White 
(2003) favor a three body encounter in which gas associated with the Taurus 
dust clouds and redshifted by the interaction with the Pleiades accounts for 
some of the red components and a separate cloud approaching from the west and 
also interacting with the cluster accounts for others. In support of the 
latter, the strongest components of CH$^+$ absorption are located northwest of 
the cluster center, with a position and orientation similar to that of the 
western cloud described by White (2003). The absence of any corresponding CH 
components in this region indicates the low density of the material. Gas 
toward 17 Tau and 23 Tau with a velocity of +10 km s$^{-1}$ likely probes the 
low-density envelope of the molecular cloud which lies just to the south of 
these sight lines. The rest of the CH$^+$ components near +9 km s$^{-1}$ may 
also trace this envelope, but the data do not preclude a connection with 
redshifted Taurus material.

The other, weaker constituent of molecular gas in the Pleiades has a velocity 
of $\sim$ +7 km s$^{-1}$, associating it with the central component of atomic 
gas and with the Taurus clouds. All sight lines with detectable amounts of CH, 
except that toward HD 23512, and the majority of CH$^+$ sight lines exhibit 
this component. It is found in both species along sight lines to three stars 
in the center of the cluster ($\eta$ Tau, 27 Tau, and 28 Tau) and these show 
similar CH/CH$^+$ column density ratios (0.2 for $\eta$ Tau and 0.1 for 27 Tau 
and 28 Tau). Toward HD 23512, where both species are detected at $\sim$ 
+9.5 km s$^{-1}$, the ratio is much larger (0.9). That the column densities of 
CH and CH$^+$ are well correlated for the central component strengthens the 
claim that CH toward most of the Pleiades is CH$^+$-like. The larger ratio 
toward HD 23512 results from the addition of CN-like CH to the total CH column 
and reinforces the idea that this sight line effectively probes the molecular 
cloud.

The patterns exhibited in the velocity components of \ion{Ca}{2} (Figure 8) 
offer unique insight into the complex interactions between the ISM and the 
stars of the Pleiades. For instance, the central component at 
$v_{\mathrm{\scriptstyle LSR}}$ $\sim$ +7 km s$^{-1}$ seen toward all of the 
stars in our survey does not have a uniform velocity across the cluster. The 
slight blueshifts and redshifts apparent in Figures 8\emph{d} and 8\emph{f}, 
respectively, are a strong indication that a dynamical interaction is 
occurring in which cloud material passing through the UV radiation field of 
the cluster is deflected toward and away from our line of sight. That the 
pattern seems to repeat itself at larger and smaller velocities (Figures 
8\emph{c} and 8\emph{g}) suggests that the encounter of this central component 
can account for at least some of the observed red and blue components. Gas 
detected at velocities higher than $\sim$ +10 km s$^{-1}$ likely derives from 
multiple sources. The feature seen in Figure 8\emph{h} may contain components 
missing from the gas in Figure 8\emph{g} which have been further redshifted by 
cluster interactions, but may also contain independent components from a 
chance meeting with a second approaching cloud. The latter is the probable 
source of the extreme-velocity components in Figure 8\emph{i} due to the 
proximity of these sight lines to the molecular cloud. The shocked atomic 
components in Figures 8\emph{a} and 8\emph{b} show additional indications of a 
velocity gradient from the center of the cluster to the northwestern edge, 
despite the noticeable gap toward 17 Tau, 20 Tau, and 22 Tau. In general, the 
various components of \ion{Ca}{2} are more pervasive in number and extent than 
those in the analysis of \ion{Na}{1} observations by White (2003), as expected 
if \ion{Ca}{2} is the most widely distributed species as in Figure 6 of Pan et 
al. (2005). Clearly, highly sensitive observations of \ion{Ca}{2} can trace 
the intricate cloud-cluster and cloud-cloud interactions of the Pleiades in 
greater detail and on a larger scale than \ion{Na}{1} observations of equal 
quality.

An important result of our analysis of \ion{Na}{1}/\ion{Ca}{2} ratios 
(\S{} 4.2) was that they predicted the proper relationship between velocity 
components and the expected physical conditions of the associated clouds, 
strengthening our confidence in the column densities derived from our 
observations. \ion{Na}{1}/\ion{Ca}{2} ratios of order a few to 10 were 
measured for most positive-velocity components indicating the diffuse nature 
of much of the interstellar gas near the cluster where calcium is not 
efficiently depleted onto the surfaces of dust grains. A ratio of order 100 
was observed in the +9.2 km s$^{-1}$ component toward HD 23512 since this gas 
lies in a denser region near the molecular cloud, an environment which results 
in higher calcium depletion. Final confirmation comes from the 
+6.2 km s$^{-1}$ component toward this star whose \ion{Na}{1}/\ion{Ca}{2} 
ratio of $\sim$ 10 rightly places it among the diffuse components.

The precise density estimates of \S{} 4.3 can place meaningful constraints on 
the chemical processes active in the diffuse clouds of the Pleiades. A 
comparison of the values obtained from CH/CH$^+$ column density ratios with 
those from rotationally excited H$_2$ column densities offers particular 
insight. We find generally good agreement in the densities obtained from the 
two methods by adopting a molecular fraction of $f=0.07$. This value of $f$ 
yields typical densities of $\sim50$ cm$^{-3}$, significantly below earlier 
predictions which favored densities between 100 cm$^{-3}$ (Gordon \& Arny 
1984) and 400 cm$^{-3}$ (White 1984\emph{a}, \emph{b}). A possible discrepancy 
with our more precise estimates comes from the recent study by Zsarg\'o and 
Federman (2003) who inferred a density upper limit of $n \leq 3.5$ cm$^{-3}$ 
from the column densities of the fine-structure levels of \ion{C}{1} toward 
$\eta$ Tau. This determination is an order of magnitude lower than our value 
toward $\eta$ Tau, $n=62$ cm$^{-3}$, obtained from both CH/CH$^+$ ratios and 
H$_2$ rotational populations. Yet, based on their density estimate, Zsarg\'o 
and Federman (2003) predict a CH column density upper limit toward $\eta$ Tau 
of $N$(CH) $\leq 7.3 \times 10^{11}$ cm$^{-2}$, consistent with our measured 
value of $4.8 \times 10^{11}$ cm$^{-2}$. To derive this predicted column 
density, the authors incorporated non-thermal motions of ions and neutrals due 
to the propagation of Alfv\'en waves in a transition zone between cloud and 
intercloud material (Federman et al. 1996\emph{a}). Without incorporating 
turbulence, the predicted column density along this sight line is $N$(CH) 
$\leq 1.6 \times 10^{9}$ cm$^{-2}$, much lower than the measured value. 
Non-thermal models hold great promise for diffuse-cloud chemistry, 
particularly in explaining the abundance of the CH$^+$ radical as well as the 
abundances of molecules like CH which are tied to its formation.

The question remains, then, why the density predicted by Zsarg\'o and Federman 
(2003) is so much lower than our value. Two possibilities immediately present 
themselves. First, observations of H$_2$, CH, and CH$^+$ may sample different 
cloud depths due to the stratification of molecular species along the line of 
sight. This scheme suggests that rotationally excited H$_2$ molecules probe 
denser portions of the cloud closer to the stars, in the main portion of the 
photodissociation region (PDR), where the UV radiation is strongest and is 
responsible for populating the higher $J$ levels through UV pumping. CH and 
CH$^+$, as well as \ion{C}{1}, would then trace the less dense outer portions 
of the cloud away from the PDR. A major drawback to this scenario is it 
requires $f=0.5$ or higher to yield low enough densities from our measured 
CH/CH$^+$ column density ratios to agree with the \ion{C}{1} result, which 
would subsequently increase our H$_2$-derived densities for this diffuse gas 
to greater than 400 cm$^{-3}$, a value more typical of the Pleiades molecular 
cloud. A more suitable alternative is that the fine-structure levels of 
\ion{C}{1} trace regions of various extent along the line of sight. If the 
$J=1$ level, for instance, was confined to a region of space 1/10 the size of 
that occupied by the $J=0$ level, the density upper limit would be more in 
line with our determination. Finally, the inclusion of time-dependent effects 
into a chemical model of CH$^+$ formation may still prove to be important 
since the timescale for H$_2$ dissociation is comparable to the characteristic 
time in which the cloud is interacting with the cluster. Because the 
initiating reaction leading to CH production requires H$_2$, a decreasing 
presence due to photodissociation would demand a higher density to account for 
the observed columns of CH. Such a requirement could be accommodated by our 
simple steady-state model by, again, raising slightly the value of $f$(H$_2$).

The most general conclusion of this investigation is that detailed 
observations of both atomic and molecular species essentially confirm the 
scenario of cloud-cluster interactions constructed by White (2003). A 
pervasive foreground gas cloud at $v_{\mathrm{\scriptstyle LSR}}$ $\sim$ 
+7 km s$^{-1}$ seen toward nearly every star in both strong atomic and weaker 
molecular absorption is flowing through the cluster and dynamically 
interacting with the stellar UV radiation field. The clearly symmetric 
patterns of both blueshifted and redshifted \ion{Ca}{2} components extending 
across the cluster demonstrate the large-scale nature of these interactions. 
A second cloud at $v_{\mathrm{\scriptstyle LSR}}$ $\sim$ +10 km s$^{-1}$ is 
also likely to be interacting with the cluster due to the strong CH$^+$ 
absorption near this velocity northwest of the cluster center and the 
existence of a diffuse molecular cloud also at $v_{\mathrm{\scriptstyle LSR}}$ 
$\sim$ +10 km s$^{-1}$ to the southwest. High velocity \ion{Ca}{2} components 
with no direct connection to redshifted foreground gas may also require the 
existence of a second interacting cloud.

\section{SUMMARY}

Twenty stars in the Pleiades were observed with high resolution and high 
signal to noise in a spectral range allowing for the detection of absorption 
features from CN, \ion{Ca}{2}~K, \ion{Ca}{1}, CH$^+$, and CH. Total equivalent 
widths are consistent with previous determinations except in cases where we 
detect weaker features owing to our greater sensitivity. Mean $b$-values for 
CN, CH$^+$, and CH indicate that toward most of the Pleiades CH is linked to 
the formation of CH$^+$ rather than to CN chemistry. Weighted mean velocities 
for individual species reveal a kinematic separation of atomic and molecular 
gas, with neutrals more closely associated with the latter. For the molecular 
species, two distinct velocity components are sufficient to characterize the 
observed profiles, with the sight line toward HD 23512 clearly probing the 
molecular cloud seen in this direction. Considerably more structure is seen 
along lines of sight in the atomic gas traced by \ion{Ca}{2}. Velocity 
gradients are apparent in the pervasive central component implying a dynamical 
interaction with the cluster. Density estimates from CH/CH$^+$ column density 
ratios agree well with those based on the rotational excitation of H$_2$ 
molecules for $f($H$_2)=0.07$, yielding typical densities of $\sim50$ 
cm$^{-3}$ for the ISM near the Pleiades. However, given the similar timescales 
of the relevant chemical and dynamical processes, the importance of 
time-dependent effects must be examined.

With the availability of high-quality absorption-line data for CN, 
\ion{Ca}{2}, \ion{Ca}{1}, CH$^+$, and CH provided by this investigation and 
for \ion{Na}{1} provided by White et al. (2001), existing models of 
interstellar chemistry must now be improved and brought to bear on the complex 
interaction between diffuse gas and the Pleiades cluster. The precise 
densities determined here must be incorporated into any model describing the 
interaction and all phenomenon must be treated self-consistently. The 
application of a time-dependent model, in particular, is likely to prove 
necessary.

\acknowledgments

We thank Yaron Sheffer for his help in synthesizing CH profiles and Dan Welty 
for his very useful comments on an earlier draft of this paper. We also 
acknowledge the thorough examination of our calculations by the anonymous 
referee that led to more consistent results. This research made use of the 
Simbad database operated at CDS, Strasbourg, France. M. Martinez participated 
in the Research Experience for Undergraduates at the University of Toledo 
under NSF-REU grants PHY-0097367 and 0353899. The research presented here was 
also supported in part by NASA Long Term Space Astrophysics grant NAG5-4957 to 
the University of Toledo.


\end{document}